\documentclass[aps,prb,reprint,floatfix,superscriptaddress]{revtex4-1}

\usepackage{amsmath,amsfonts,amssymb} %
\usepackage{mathtools} %
\usepackage{wasysym} %
\usepackage{bm} %
\usepackage[bbgreekl]{mathbbol} %
\usepackage{graphicx} %
\usepackage[acronym]{glossaries} %
\usepackage{braket} %
\newcommand{\Fig}[1]{Figure~(\ref{#1})} %
\newcommand{\eq}[1]{Eq.~(\ref{#1})} %
\newcommand{\fig}[1]{Fig.~\ref{#1}} %
\def\be{\begin{equation}} %
\def\ee{\end{equation}} %

\newcommand{\bea}{\begin{eqnarray}}
\newcommand{\eea}{\end{eqnarray}}

\newcommand{\BRA}[1]{\langle{} #1 \vert}
\newcommand{\KET}[1]{\vert{} #1 \rangle}

\newcommand{\DE}[1]{\Delta \epsilon_{#1}}
\newcommand{\DR}[1]{\Delta \rho_{#1}}

\begin{document}

\title{Coherent control based on quantum Zeno and anti-Zeno effects: Role of coherences and timing}

  \author{Jacob Levitt}
\affiliation{Cortex Fusion Systems, Inc., New York, NY, 10128, United States}
\email{jakelevitt@cortexfusion.systems}

\author{Artur F. Izmaylov}
\email{artur.izmaylov@utoronto.ca}
\affiliation{Department of Physical and Environmental Sciences,
  University of Toronto Scarborough, Toronto, Ontario, M1C 1A4,
  Canada}
\affiliation{Chemical Physics Theory Group, Department of Chemistry,
  University of Toronto, Toronto, Ontario, M5S 3H6, Canada}

\date{\today}
\begin{abstract}

The quantum-Zeno and anti-Zeno effects (QZE/AZE) 
are known for a long time, in a quantum system with coupled levels, the measurement 
of a particular level population can lead to 
either acceleration (i.e. AZE) or retardation (i.e. QZE) of its population transfer to other levels.    
Here we consider how one can control the population flow from a coupled quantum state by measurement at a particular time, and what system parameters are responsible for that.
We propose a framework for analysis of quantum Zeno dynamics based on time-dependent density matrix perturbation theory. This framework allows us to clearly separate state populations from their coherences
and to predict appearance of either QZE or AZE.  
We illustrate our analysis on two model systems: 
1) two coupled levels and 2) a level coupled to a continuum. In both cases dynamics of quantum 
coherences play a crucial role, and perturbative considerations allow us to predict the effect of 
projective measurements. In addition, we have extended our consideration to a closely related coherent 
control scenario, a unitary transformation altering the sign of a state in a coherent superposition describing the system wavefunction.  

\end{abstract}
%\date{\today}
\maketitle

\section{Introduction}

Exploiting various quantum mechanical effects (e.g. quantum coherences and entanglement) comprises a key development in the arrival of quantum technology into modern society\cite{Milburn1997-yb}. 
One prodigious insight into quantum control is the quantum-Zeno effect (QZE) \cite{misra1977zeno,itano1990quantum,kofman1996quantum,home1997conceptual,koshino2005quantum,thapliyal2016linear}, for which the quantum system can be prevented from sampling certain subspaces of its Hilbert space by applying the appropriate series of projective measurements. Extensions of the QZE to include non-demolition measurements \cite{gagen1992quantum} and unitary operators \cite{facchi2004unification,dhar2005preserving}, as well as the complementary, anti-Zeno effect (AZE), in which the quantum system is forced to sample its own Hilbert space in any arbitrary desired fashion \cite{koshino2005quantum,thapliyal2016linear,kofman2000acceleration,facchi2001quantum,yamaguchi2008photon}, all fall under the umbrella of the QZE mechanism \cite{fischer2001observation,facchi2002quantum,schafer2014experimental,signoles2014confined,gherardini2016stochastic,muller2016stochastic,Facchi2008,Kwiat1999,Zheng2008}. The QZE and AZE have demonstrated diverse utility in key technology areas such as quantum sensing \cite{virzi2022quantum,do2019experimental,long2022entanglement,muller2016stochastic}, chiral spectroscopy \cite{wu2020discrimination}, integrated photonics \cite{liu2023engineering,thapliyal2016linear,longhi2006nonexponential,das2021interplay,das2023quantum,zezyulin2012macroscopic,huang2010interaction}, quantum communication \cite{nodurft2022polarization,nodurft2022generation,jacobs2009all,huang2010interaction,zhou2009quantum,hendrickson2013all,nodurft2022entangling,ten2011entangling,bayrakci2022quantum,tavakoli2015quantum,cao2017direct}, quantum counterfactual communication \cite{salih2022laws,cao2017direct,salih2013protocol,vaidman2019analysis,cao2014direct,alonso2019trace,li2015direct}, Bell-state analysis \cite{zaman2018counterfactual}, non-demolition measurement \cite{ma2014chip,leppenen2021quantum,helmer2007quantum}, and quantum computing \cite{maniscalco2008protecting,HacohenGourgy2018,PazSilva2012,Wang2008,Kofman2004,Harrington2017,franson2004quantum}. They are generally considered paradigmatic for quantum control \cite{Mller2017,Kofman2001,zhou2009quantum,HacohenGourgy2018,Pechen2006,Srensen2018,maniscalco2008protecting,Piacentini2017,PazSilva2012,Wang2008,Barontini2015,Kofman2001b,Kofman2004,Clausen2010,Harrington2017}: one goal of the present work is to deliver a unified framework of the QZE/AZE with the aspiration that this enables its further penetration into quantum science applications.

Even though both QZE and AZE were modelled and exploited for long time, clear criteria on when one or the 
other occurs in a particular system have not been fully established. Kofman and Kurizki in their seminal work (2000) \cite{kofman2000acceleration} have shown that both effects can occur in a level coupled to a 
continuum system. A particular dynamical scenario after the measurement 
is determined by the overlap of the level broadening (due to the coupling with the continuum) and the continuum spectral density.\cite{kofman2000acceleration,PhysRevA.98.012135}
This generally corresponds to observing QZE(AZE) if the level is energetically within (outside of) the range of 
the continuum energies.  
Yet, there are other model systems, some of them we are considering below, where QZE and AZE can 
be present in the same system and a switch between them is determined by the time of the measurement.
 
One clarification on terminology needs to be done here, what is meant by the measurement process in the 
quantum Zeno dynamics literature refers only to decoherence by the device and not the selection of 
the pointer state. In other words, the measurement is assumed to produce the mixed state 
$\rho_M = \sum_i |c_i|^2 \ket{i}\bra{i}$ from the initial pure state of the evolving quantum system, 
$\rho = \ket{\psi}\bra{\psi} = \sum_{ij} c_i^{*}c_j \ket{i}\bra{j}$ due to interaction with the measuring device. 
In this case, the measurement can be seen as enforcing unitary dynamics of the system coupled to the 
device, which is equivalent to the von Neumann measurement.\cite{vonNeumann1955} 

Considering that QZE and AZE are related to alternation of coherences by the measurement, 
one can attempt to control the system by applying unitary transformations that 
only change signs of its coherences instead of destroying them.        
Saha {\it et al.} (2011) \cite{saha2011tunneling} applied a perturbative framework to explore 
the QZE and AZE for scenarios involving both projective (demolition) and unitary (non-demolition) interactions with a quantum system. Population transfer between two weakly coupled states, where the target state is embedded into a continuum, was shown to be somewhat controlled via modification the timing between successive measurements or unitary rotations that change signs of coherences between system states. 
In Ref.~\citenum{saha2012tunneling}, this result was extended to the problem of nuclear fusion via a model of tunnelling through coulombic (MeV-scale) barriers providing some evidence that unitary control of a bound nuclear wavepacket can accelerate population transfer to the continuum of fusion product channels.
Yet, there was no attempt to elucidate what is responsible for appearance of a particular effect, QZE or AZE.  

In this work we would like to clarify the role of quantum coherences that appear during the quantum 
dynamics associated with the population transfer. Clearly, these coherences are involved in control 
scenario involving QZE, AZE, or their analogues stimulated by unitary transformations that change 
system state signs in the system superposition wavefunction. To perform this analysis we will employ 
time-dependent density matrix perturbation theory, where the perturbation is responsible for coupling 
between the system states.   

\section{Theory}

\subsection{Two-level model}
Let us consider a two-level model given by the Hamiltonian 
\bea
H = \epsilon_0\ket{0}\bra{0}+\epsilon_1\ket{1}\bra{1}+V(\ket{1}\bra{0}+\ket{0}\bra{1}),
\eea
where $\epsilon_i$ are level energies, and $V$ is a coupling. The density matrix of the system is 
$\rho(t) = \sum_{i,j} \rho_{ij}(t) \ket{i}\bra{j}$, satisfying the Liuovillian equation
$i\partial_t \rho(t) = [H, \rho(t)]$ with the initial condition $\rho(0) = \ket{0}\bra{0}$ 
and the population of the $0^{\rm th}$ state, $\rho_{00}$, as the quantity of interest. 
All numerical simulations are done with the two-state model with $V=0.2$ a.u., 
$\epsilon_{1/0} = \pm 0.2$ a.u. Here and in what follows we will use atomic units.

To see the QZE and AZE from measurements and unitary transformations, it is 
convenient to consider perturbation theory analysis of $\rho(t)$ dynamics treating 
$\hat V = V(\ket{1}\bra{0}+\ket{0}\bra{1})$ as the perturbation. In the interaction picture, 
$\rho_I(t) = e^{iH_0t}\rho(t)e^{-iH_0t}$, up to the second order in perturbation,
the density is given as  
\bea
\rho_I(t) &=& \rho(0)+\rho_I^{(1)}(t) +\rho_I^{(2)}(t), \\
\rho_I^{(1)}(t) &=& -i\int_0^{t} d\tau [V_I(\tau),\rho(0)], \\
\rho_I^{(2)}(t) &=& -\int_0^{t}d\tau \int_0^{\tau} d\tau' [V_I(\tau),[V_I(\tau'),\rho(0)]],
\eea
where $V_I(\tau) = e^{-iH_0\tau} \hat V e^{iH_0\tau}$. 
%change in the total density from time $0$ to time $t$ is $\rho_I(t)$
The first order in $V_I$ process produces populations $\rho_{ss}^{(1)}(t)$ from coherences 
$\rho_{ss'}(0)$
\bea
 \rho_{ss}^{(1)}(t) &=& -i \BRA{s} \int_0^{t} d\tau [V_I(\tau),\rho(0)]\KET{s}\\ \notag
 &=&-i \int_0^{t} d\tau\BRA{s}V_I(\tau)\KET{s'}\rho_{s's}(0) \\
 &&-\rho_{ss'}(0)\BRA{s'}V_I(\tau)\KET{s} \\
 &=&2{\rm Im}\left[ \rho_{s's}(0) \int_0^{t} d\tau\BRA{s}V_I(\tau)\KET{s'}\right] \label{eq:CPX}
 \eea
and coherences $\rho_{ss'}^{(1)}(t)$ from populations $\rho_{ss}(0)$
\bea
 \rho_{ss'}^{(1)}(t) &=& -i \BRA{s} \int_0^{t} d\tau [V_I(\tau),\rho(0)]\KET{s'}\\
 &=&i[\rho_{ss}(0)-\rho_{s's'}(0)] \int_0^{t} d\tau\BRA{s}V_I(\tau)\KET{s'},\label{eq:PC1}
 \eea
the last equality is the consequence of zero diagonal perturbation terms, 
$\BRA{s}V_I(\tau)\KET{s} = 0$. 
In the second order, due to the absence of the diagonal perturbation elements, there is no contribution 
from coherences to population dynamics, the only contribution originates from populations  
 \bea\nonumber
\rho_{ss}^{(2)}(t) &=& \BRA{s}\int_0^{t}d\tau \int_0^{\tau} d\tau' [V_I(\tau),[V_I(\tau'),\rho(0)]]\KET{s} \\
&=& \int_{0}^{t} d\tau\int_{0}^{t} d\tau' \rho_{ss}(0)\BRA{s} V_I(\tau) V_I(\tau')\KET{s}  \\ \notag%\\ \label{eq:ps2x}
&&-\BRA{s} V_I(\tau) \KET{s'}\rho_{s's'}(0)\BRA{s'} V_I(\tau')\KET{s} \\
&=& [\rho_{ss}(0)-\rho_{s's'}(0)] \\ \notag
&&\times\int_{0}^{t} d\tau\int_{0}^{t} d\tau' \BRA{s} V_I(\tau) \KET{s'}\BRA{s'} V_I(\tau')\KET{s}.\label{eq:PP2}
\eea

% Coherence to coherence in the 2nd order - useful for U when coherences are not zeroed.

To follow populations $\rho_{ss}(t)$ we need to consider two paths: 1) two first order steps, 
population to coherences [\eq{eq:PC1}] and coherences to populations [\eq{eq:CPX}], 
and 2) a single second order process, population to population [\eq{eq:PP2}]. 
For small enough times higher order processes can be ignored 
because no matter how large $V$ is, it is multiplied by time in all equations for final quantities. 

\begin{figure}[h!]
    \centering
    \includegraphics[width=1.0\linewidth]{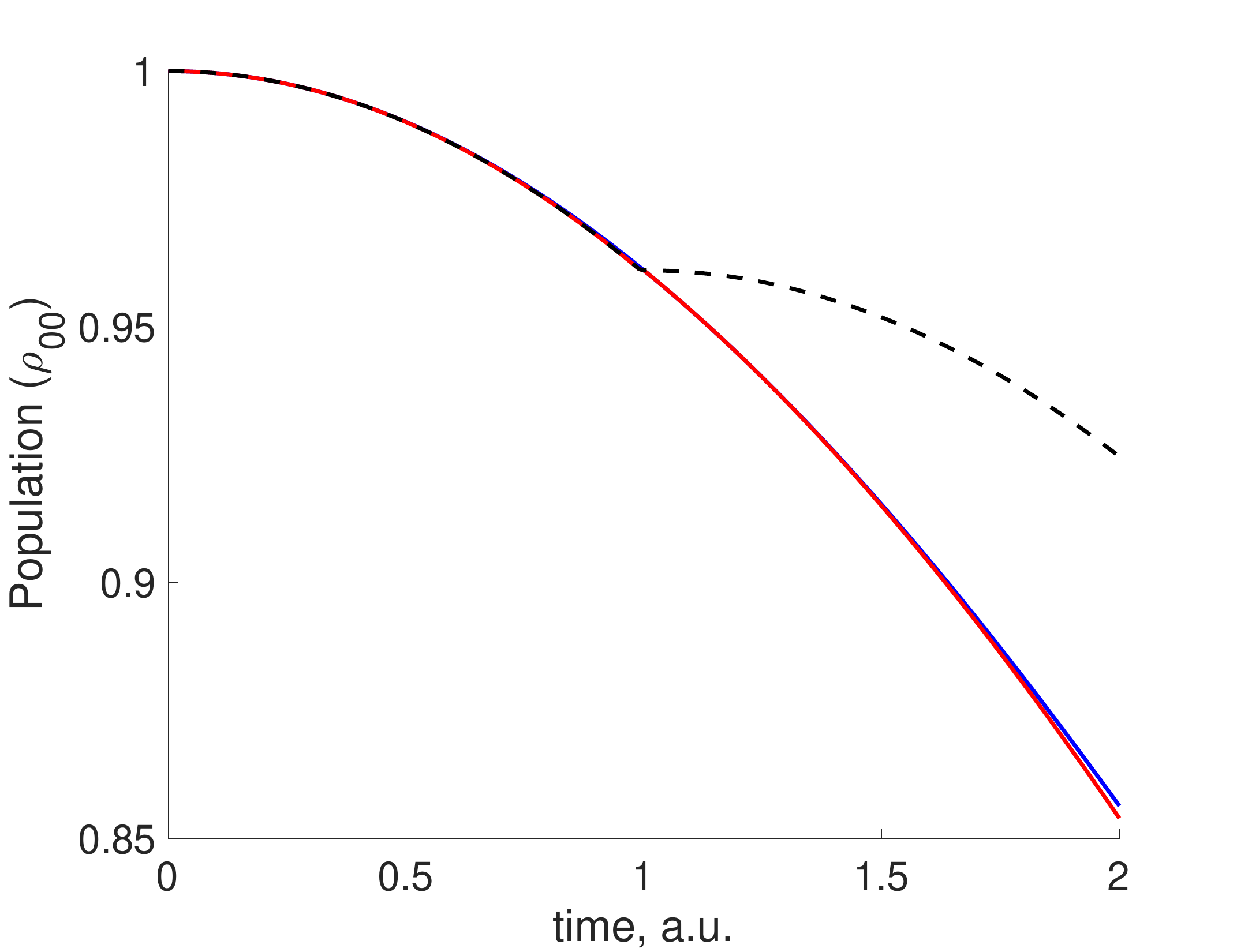}
    \caption{Population $\rho_{00}$ given by perturbation theory (red solid), exact (blue solid), 
    and exact dynamics with coherences destroyed due to a measurement at 1 a.u. (black dashed).}
    \label{fig:Zeno}
\end{figure}

The QZE can be explained by noticing that if measurement destroys coherences,
$\rho_{ss'}\rightarrow 0$, then the population transfer path 1 is disrupted because \eq{eq:CPX}
will have zero coherences at the moment after the measurement. 
This disruption leaves only path 2 that has a quadratic dependence 
on time, $\rho_{ss}(t)\sim t^2$, which has a zero slope at $t=0$ and the time after the measurement 
(see \fig{fig:Zeno}, here we assume that the measurement is instantaneous). 

\begin{figure}[h!]
    \centering
    \includegraphics[width=1.0\linewidth]{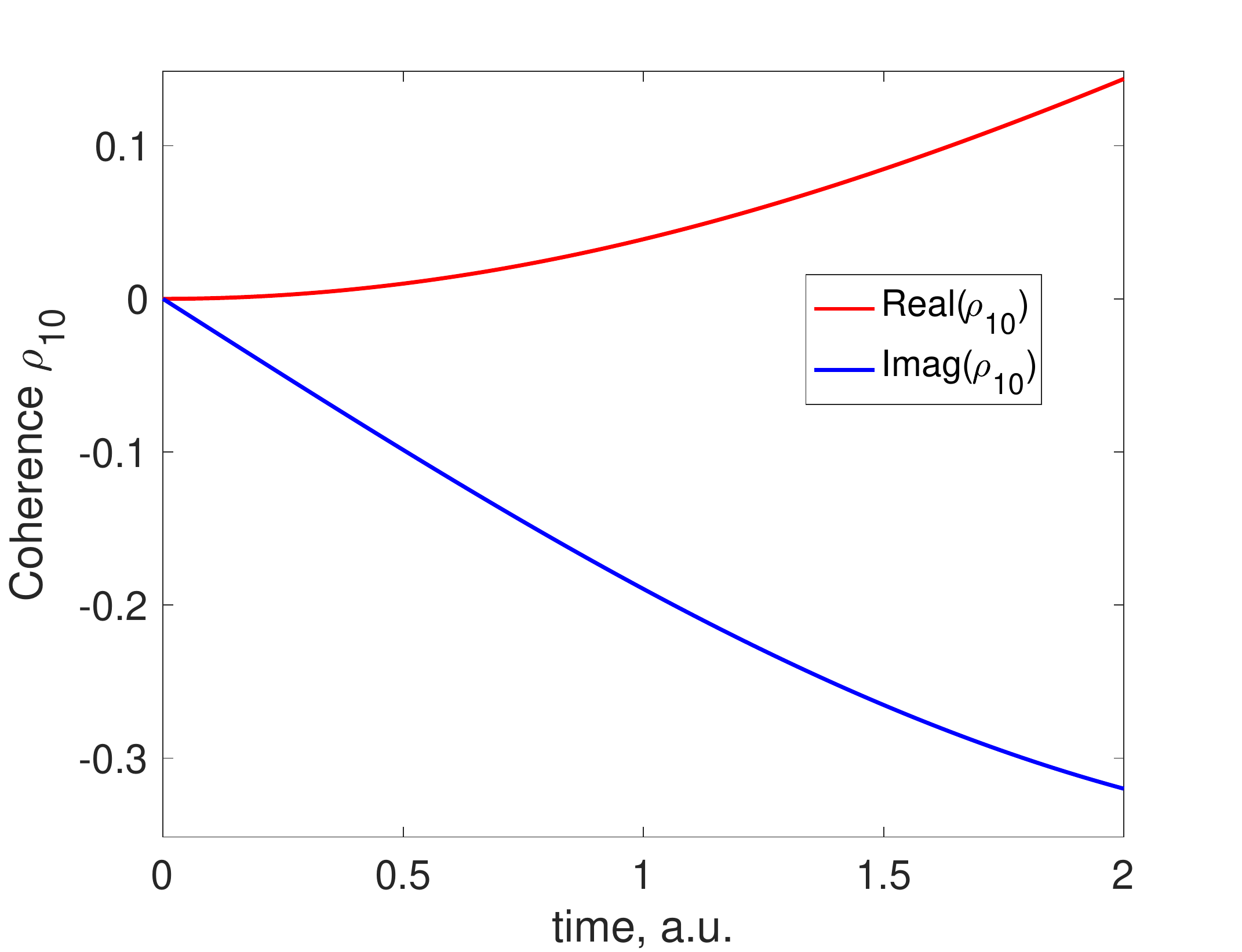}
    \caption{Coherence $\rho_{10}$ real and imaginary components during the exact dynamics. 
    Negative values of the imaginary part are responsible for positive contribution to the overall population 
    transfer from $0$ to $1$. }
    \label{fig:ZenoC}
\end{figure}

\begin{figure}[h!]
    \centering
    \includegraphics[width=1.0\linewidth]{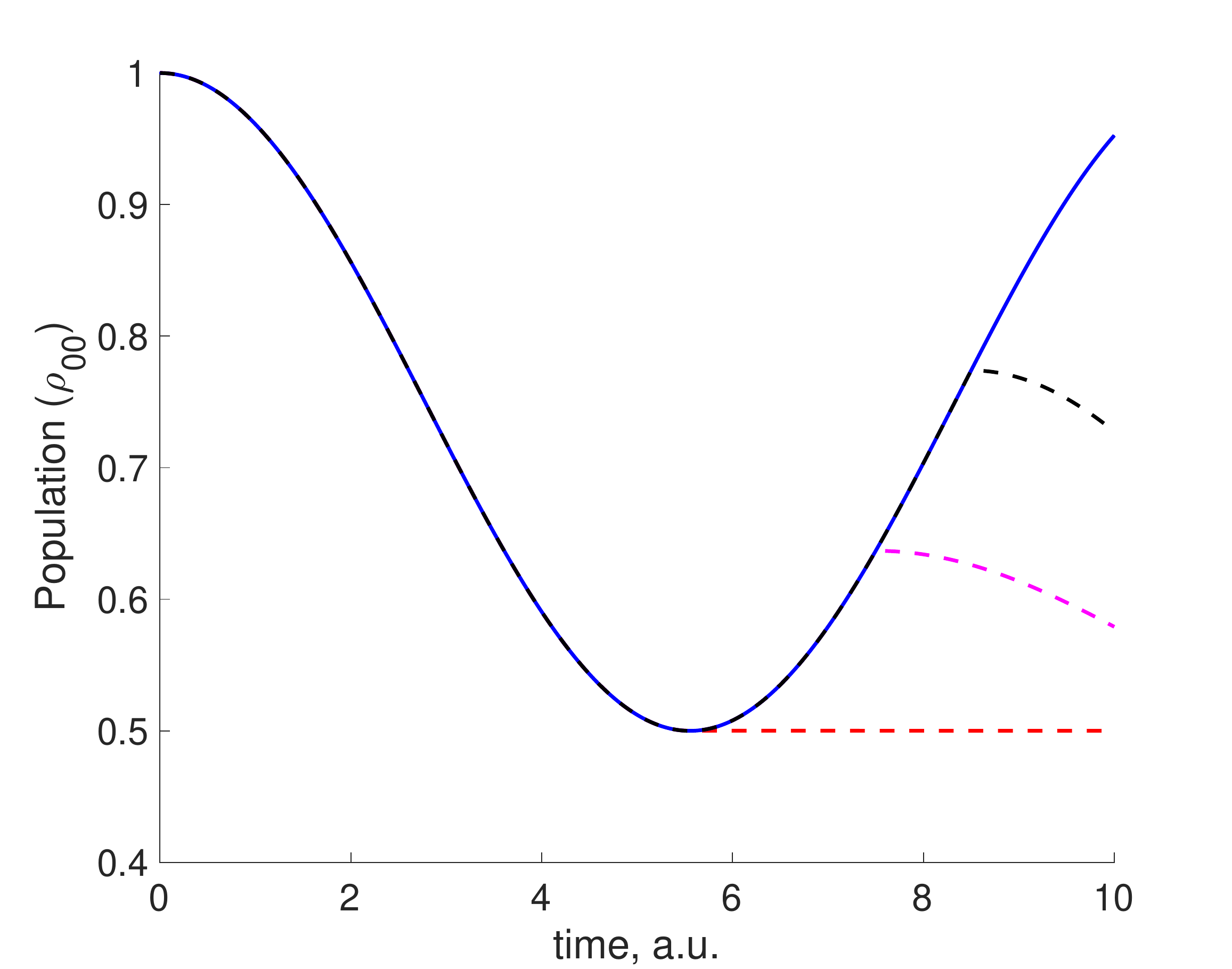}
    \caption{Population $\rho_{00}$ given by exact dynamics (blue solid) and three measurement induced coherent destruction dynamics that are different in the time of measurement: 5.5 a.u. (black dashed), 7.5 a.u. 
    (magenta dashed), and 8.5 a.u. (blue dashed).}
    \label{fig:aZeno}
\end{figure}

\begin{figure}[h!]
    \centering
    \includegraphics[width=1.0\linewidth]{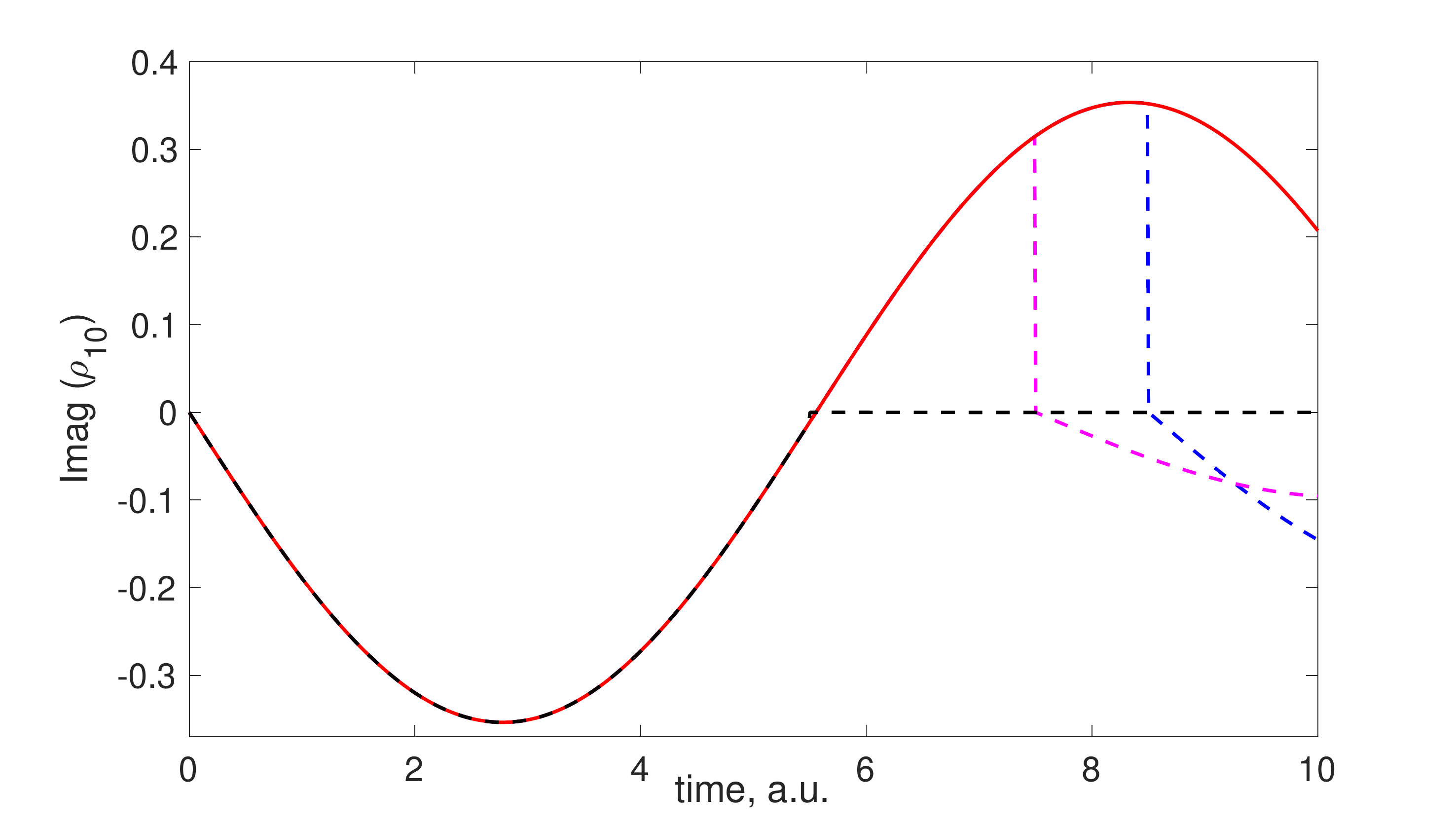}
    \caption{Imaginary part of $\rho_{10}$ given by exact dynamics (red solid) and three measurement induced coherence destruction dynamics that are different in the time of measurement: 5.5 a.u. (red dashed), 7.5 a.u. 
    (magenta dashed), and 8.5 a.u. (black dashed).}
    \label{fig:aZenoC}
\end{figure}

For understanding the AZE, note that coherences can enhance 
both forward ($\rho_{00}\rightarrow\rho_{11}$) and backward  ($\rho_{11}\rightarrow\rho_{00}$) transfers
according to \eq{eq:CPX}. Figure \ref{fig:ZenoC} illustrates that during the forward process negative imaginary part of $\rho_{10}$ coherence builds up and facilitates the transfer via path 1. 
Due to the sum rule $\sum_s \rho_{ss}=1$, disruption  of the 
backward transfer is enhancing the forward transfer. Therefore, if the coherences are destroyed by the 
measurement at the time when their values stimulate backward transfer one will observe the 
AZE (see \fig{fig:aZeno}). There are few considerations that help understanding these dynamics. 
First, due to the level energy difference the transfer is incomplete state $\ket{1}$ is only populated 50\%, this takes place around 5.5 a.u. This makes population difference $\rho_{00}-\rho_{11}$ always positive and thus path 2 will always favors the forward flow of the population, 
$\rho_{00}\rightarrow\rho_{11}$. Second, \fig{fig:aZenoC} shows that the imaginary part of coherence $\rho_{10}$ becomes positive after 5.5 a.u., which makes the coherence to facilitate the reverse process, 
$\rho_{11}\rightarrow\rho_{00}$,  according to path 1. Considering these two points, the presence of the AZE on 
\fig{fig:aZeno} becomes clear. Destroying the coherences around 5.5 a.u. leaves only the population 
channel (path 2) with $\rho_{00}-\rho_{11}\approx 0$ which does not produce either good population transfer 
or build up of new coherences. If one does the measurement later, 7.5 a.u. or 8.5. a.u., there is a larger population difference $\rho_{00}-\rho_{11}$, but this difference still favours the forward transfer. Also, when 
the coherence $\rho_{10}$ is destroyed and it starts to build up again, its imaginary part has a different sign 
than before the measurement (\fig{fig:aZenoC}), hence, 
this also favours the forward path. 

\begin{figure}[h!]
    \centering
    \includegraphics[width=1.0\linewidth]{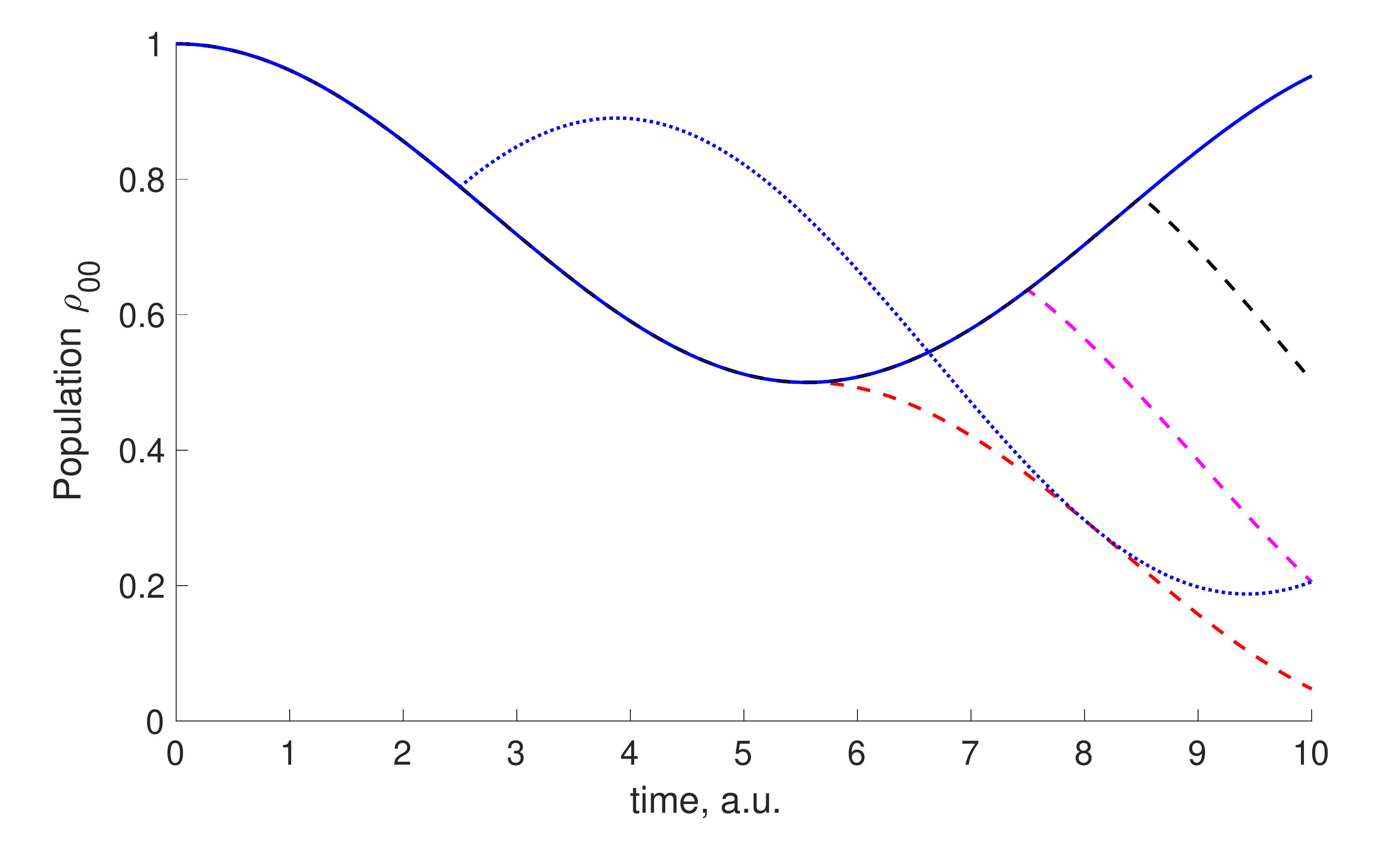}
    \caption{Population $\rho_{00}$ given by exact dynamics (blue solid) and four $U$ induced coherence 
    sign-flip dynamics that are different in the time of $U$ application: 2.5 a.u. (blue dotted), 5.5 a.u. (black dashed), 7.5 a.u. (magenta dashed), and 8.5 a.u. (blue dashed).}
    \label{fig:UZeno}
\end{figure}

Moreover, one can see from \eq{eq:CPX} that the direction 
in which coherences sway the population flow depend on the sign of their imaginary part. Thus,
if one can change only the sign of coherences it will be enough to change the direction of the 
population flow from \eq{eq:CPX}. This sign change can be accomplished by the unitary 
transformation $U = 1- 2\ket{0}\bra{0}$, which is also hermitian $U^\dagger = U$. 
$U$ is changing the sign of the coherences without changing populations
\bea\notag
U^\dagger \rho U &=& \rho_{00} \ket{0}\bra{0}+\rho_{11} \ket{1}\bra{1} \\
&&- \rho_{01} \ket{0}\bra{1} - \rho_{10} \ket{1}\bra{0}. 
\eea
Figure \ref{fig:UZeno} shows how $U$ changes the population transfer if applied in different times. 
When applied before the imaginary part of the $\rho_{10}$ coherence switches from negative to positive 
($t<5.5$ a.u. according to \fig{fig:aZenoC}) the $U$ sign-flipping effect will be similar to the QZE. 
However, if $U$ is applied when ${\rm Im}[\rho_{10}]>0$, $t>5.5$ a.u., it enhances the 
AZE to the point where more forward population transfer becomes possible. 
Note that because $U$ flips the sign of non-eigenstate coherences, it can change the system 
energy.     

To avoid dependence on limitations of perturbation theory at longer times, we continue here with a parallel non-perturbative consideration, which provides the following equations of motion:
\bea
i\partial_t\rho_{jj} &=& \bra{j} [H,\rho] \ket{j} = \sum_k H_{jk}\rho_{kj} - \rho_{jk}H_{kj} \\
i\partial_t\rho_{jk} &=& \bra{j} [H,\rho] \ket{k} = \sum_l H_{jl}\rho_{lk} - \rho_{jl}H_{lk} \\
&=& \sum_l H_{lj}\rho_{lk} - \rho_{lj}^{*}H_{lk} \\
&=& (\epsilon_j-\epsilon_k)\rho_{jk} - 
(\rho_{jj}-\rho_{kk}) V,
\eea
where we used real character of our Hamiltonian matrix elements $H_{jk} = H_{kj}$ and hermiticity of 
the density matrix $\rho_{jk} = \rho_{kj}^{*}$. 
Considering that dynamics is unitary and populations always stay real gives
\bea\label{eq:SumImC}
\partial_t {\rm Re} [\rho_{jj}] &=& \sum_{k\ne j}  2 H_{jk} {~\rm Im}(\rho_{kj}),  
\eea
this shows importance of signs and magnitudes of the imaginary parts 
of coherences for the population dynamics. For the 2-level system this 
provides $\partial_t {\rm Re} [\rho_{00}] = 2 V{~\rm Im}(\rho_{01})$.

It is also useful to have the expressions for time derivatives of 
real and imaginary parts of coherences
\bea
\partial_t {\rm Re}(\rho_{jk}) &=& (\epsilon_j-\epsilon_k){\rm Im}(\rho_{jk}) \\ \label{eq:ImCtd}
\partial_t {\rm Im}(\rho_{jk}) &=& (\rho_{jj}-\rho_{kk})V -(\epsilon_j-\epsilon_k){\rm Re}(\rho_{jk}). 
\eea
Since both QZE and AZE heavily depend on coherence dynamics it is worth emphasizing that 
a significant difference between dynamics of the system after the measurement and after the $U$ 
transformation is the presence of the ${\rm Re}(\rho_{jk})$ term [\eq{eq:ImCtd}]. 
The measurement destroys 
coherences and make the ${\rm Re}(\rho_{jk})$ term zero, while $U$ only inverts its sign. The implications of this is 
already visible by comparing \fig{fig:aZeno} and \fig{fig:UZeno}, the latter has much more 
pronounced AZE at 5.5 a.u.

\subsection{Level coupled to a continuum}

The Hamiltonian for this type of systems is 
\bea
H = \epsilon_0\ket{0}\bra{0}+ \sum_k\epsilon_k\ket{k}\bra{k}+V(\ket{k}\bra{0}+\ket{0}\bra{k}),\quad
\eea
where $V=0.01$, 
$\epsilon_k \in [-D,D]$,  $D=5$ a.u., $\epsilon_0 = 0$ for the level in a continuum (LIC) model 
and $\epsilon_0=D+0.04$ for the level outside of a continuum (LOC) model. In our simulations, 
the continuum is modelled by 200 discrete levels uniformly distributed in the $[-D,D]$ range  
so that $\epsilon_{k+1}-\epsilon_{k}= 0.05$ a.u.  

\begin{figure}[h!]
    \centering
    \includegraphics[width=1.0\linewidth]{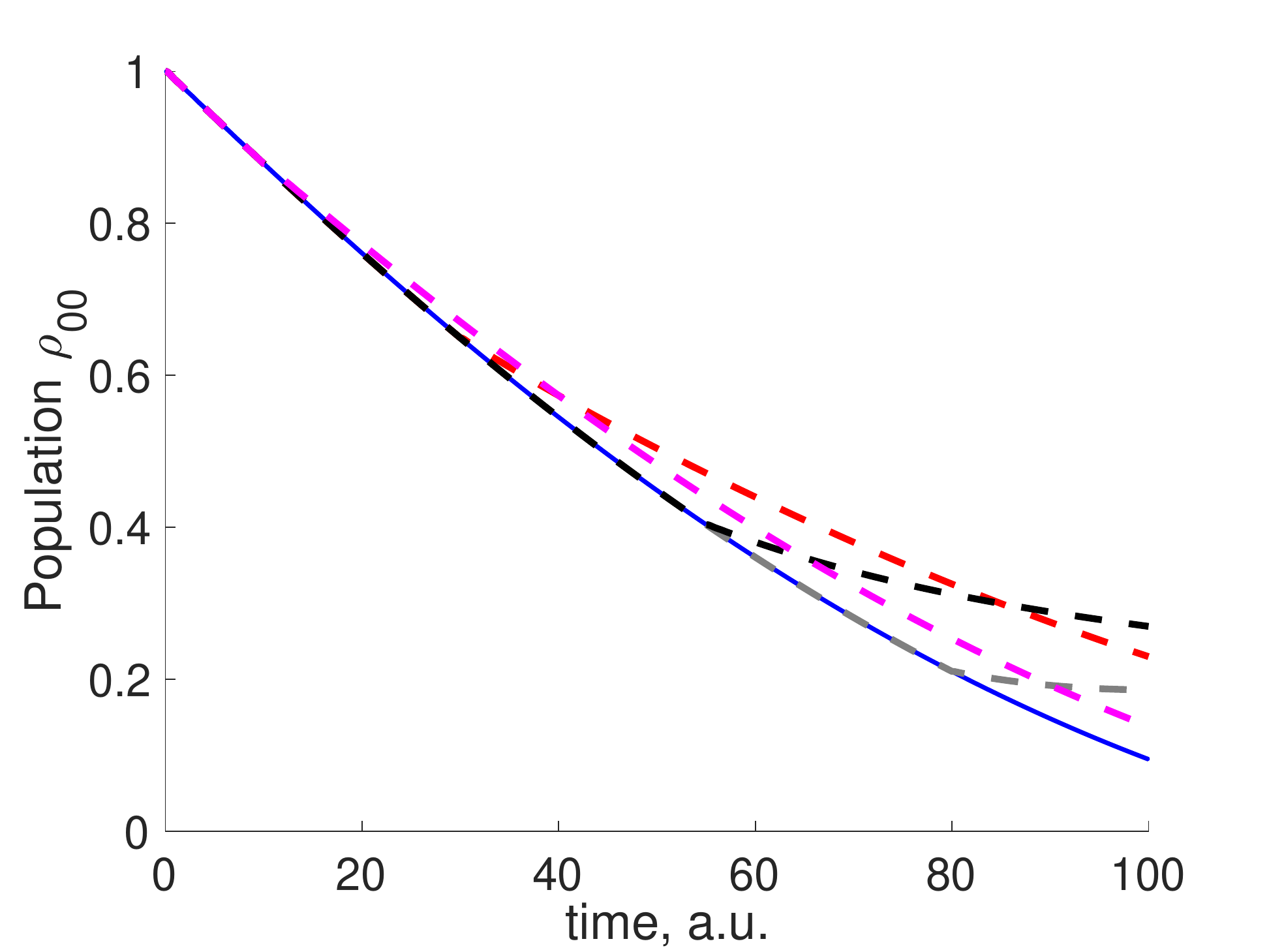}
    \caption{Population $\rho_{00}$ given by exact dynamics (blue solid) and four measurement 
    induced decoherences that are done in different times: 10 a.u. (magenta dashed), 30 a.u. (red dashed), 55 a.u. (black dashed), and 80 a.u. (grey dashed).}
    \label{fig:LCZeno}
\end{figure}

\begin{figure}[h!]
    \centering
    \includegraphics[width=1.0\linewidth]{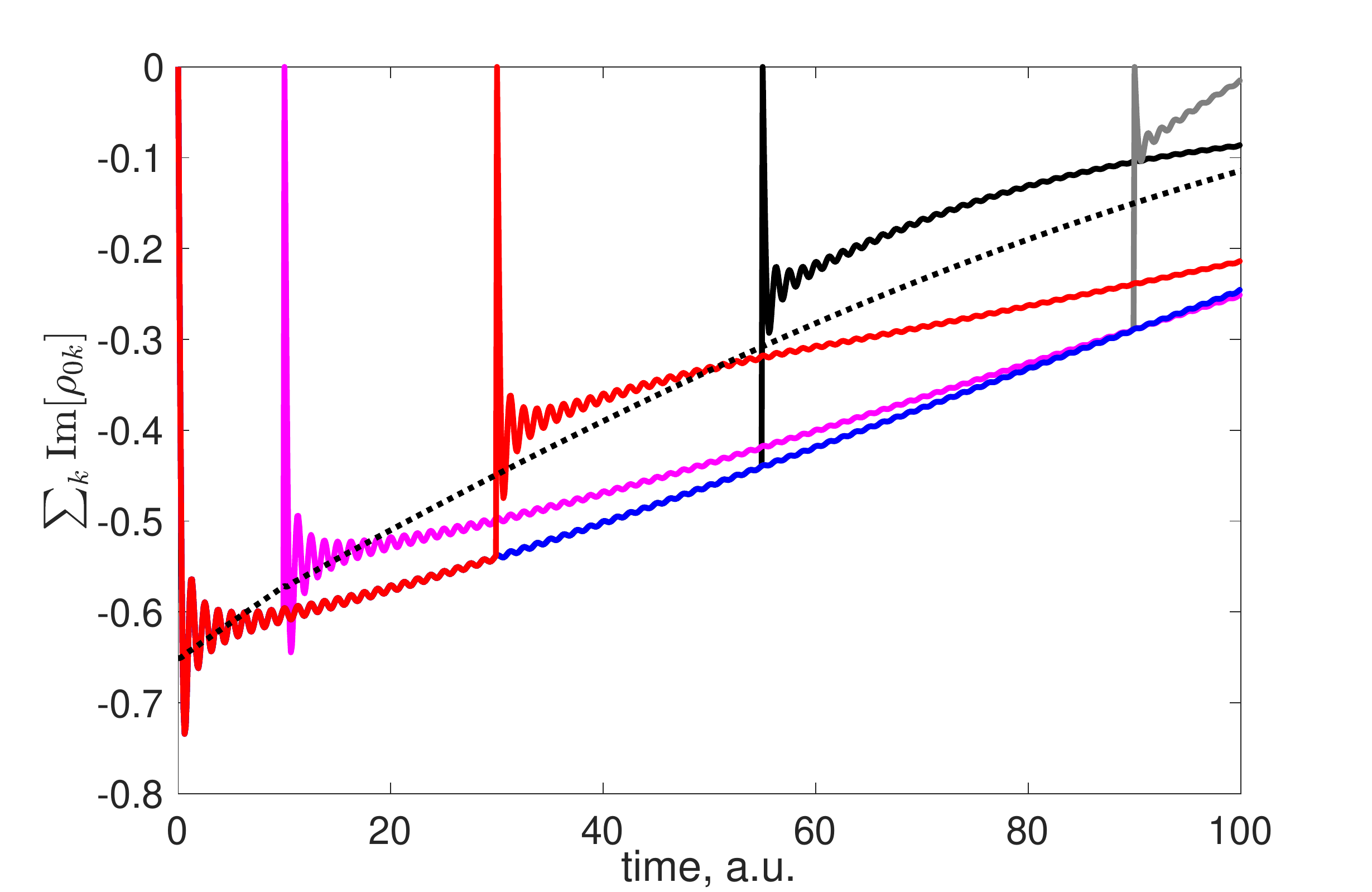}
    \caption{Sum of imaginary part of coherences between the level and continuum states 
    ($\Sigma(t)$) as a function of time, given by exact dynamics (blue solid) and four measurement induced decoherences that are done in different times: 10 a.u. (magenta dashed), 30 a.u. (red dashed), 55 a.u. (black dashed), and 90 a.u. (grey dashed). The dotted black line is $\Sigma^{(1)}(t_{\min})$ from \eq{eq:St_m}, its time dependence comes from $\DR{k}$ evaluated along the exact dynamics.}
    \label{fig:LICCZ}
\end{figure}

\paragraph{The level in a continuum model:} Exact dynamics starting at the $\ket{0}$ state gives the 
population decay to the continuum (\fig{fig:LCZeno}). As expected, there is an exponential decay 
of the population which has a recurrence at later time due to a finite continuum discretization (the recurrence 
is not shown on the figure). When at various times we perform measurements collapsing all 
coherences that are created, the overall trend is the QZE, slowing down the transfer 
from the $\ket{0}$ state to the continuum. The questions that arise are what explains the QZE 
here and why it is getting stronger at later time. For both questions, it is instructive to 
consider a total imaginary part of all coherences between the level and the continuum states, 
$\Sigma = \sum_k {\rm Im} [\rho_{0k}]$ (\fig{fig:LICCZ}). According to \eq{eq:SumImC}, 
$\Sigma$ is the quantity that determines the coherence channel of the population transfer. 
\Fig{fig:LICCZ} reveals a few trends in $\Sigma$ dynamics. Initially, $\Sigma$ quickly becomes negative via oscillations that also quickly reduce the amplitude and turn into a stable growth of 
$\Sigma$. This $\Sigma(t)$ dynamics can be easily understood from the perturbation theory point of view
for the first order channel from the populations to coherences, \eq{eq:PC1}. 
Integrating the time and taking imaginary part in \eq{eq:PC1} for our model gives 
\bea\label{eq:SigmaT}
\Sigma^{(1)}(t) = V \sum_k \DR{k}\frac{\sin(\DE{k}t)}{\DE{k}},
\eea  
where $\DR{k} = \rho_{kk} - \rho_{00}$ and $\DE{k} = \epsilon_{k} - \epsilon_0$. Clearly, 
all $\DR{k}$ are $-1$ at the initial time, and sum of $\sin(\DE{k}t)/\DE{k}$ stabilizes very quickly 
due to cancellations of oscillations from different frequencies $\DE{k}$. 
The overall growth of $\Sigma(t)\approx\Sigma^{(1)}(t)$ comes from 
a time-dependence of $\DR{k}$ that is assumed to be negligible in the first order of 
perturbation theory.   

The negative values of $\Sigma$ along the dynamics stimulates forward population flow 
$\rho_{00}\rightarrow \rho_{kk}$ via coherences (\eq{eq:CPX} or \eq{eq:SumImC}). Because 
of $\Sigma(t)$ growth with time, this channel becomes weaker. To understand the QZE, 
we only need to notice that after the disruption of the coherence channel $\Sigma(t)$
returns to higher values than before the measurement, thus the coherent channel for the 
population transfer is weakened by the measurement. \Fig{fig:LICCZ} shows this clearly 
if we compare the $\Sigma(t)$ level dips right after the measurements and the 
$\Sigma(t)$ level in dynamics without measurements. Yet, this still leaves a question 
why $\Sigma(t)$ plunges right after the measurement?

The dip of $\Sigma(t)$ right after the measurement can be well described 
using the perturbation theory from \eq{eq:SigmaT}. 
Considering that only short time is required to reach the first minimum of 
$\Sigma(t)$ in $t$, one can easily find 
approximations to $t_{\min}$ and $\Sigma(t_{\min})$ using a Taylor decomposition for 
$\sin(\DE{k}t)\approx \DE{k}t - \DE{k}^3t^3/6$ in $\Sigma^{(1)}(t)$,  
\bea
t_{\min} &=& \sqrt{2\frac{\vert\sum_k \DR{k}\vert}
{\vert\sum_k \DR{k}\DE{k}^2\vert}}\\ \label{eq:St_m}
\Sigma^{(1)}(t_{\min}) &=& -\frac{V}{3} \frac{\vert 2\sum_k \DR{k}\vert^{3/2}}
{\vert\sum_k \DR{k}\DE{k}^2\vert^{1/2}}.
\eea
\Fig{fig:LICCZ} shows that $\Sigma^{(1)}(t_{\min})$ approximates well the first minima right 
after each measurement in $\Sigma(t)$. 
Every time the measurement is done, $\Sigma(t)$ vanishes first and then quickly returns to values well approximated by $\Sigma^{(1)}(t_{\min})$ 
but with different values of  $\DR{k}$ that evolve in time (\fig{fig:LICCZ}). When we calculate 
$\Sigma^{(1)}(t_{\min})$ we use time-dependent $\DR{k}$ from exact quantum dynamics. 
To summarize, the QZE in this case occurs from weakening the coherence channel 
by zeroing the coherences, this leads to a higher $\Sigma(t)$ level due to higher values of $\DR{k}$ than 
at the starting point of the dynamics.     

\begin{figure}[h!]
    \centering
    \includegraphics[width=1.0\linewidth]{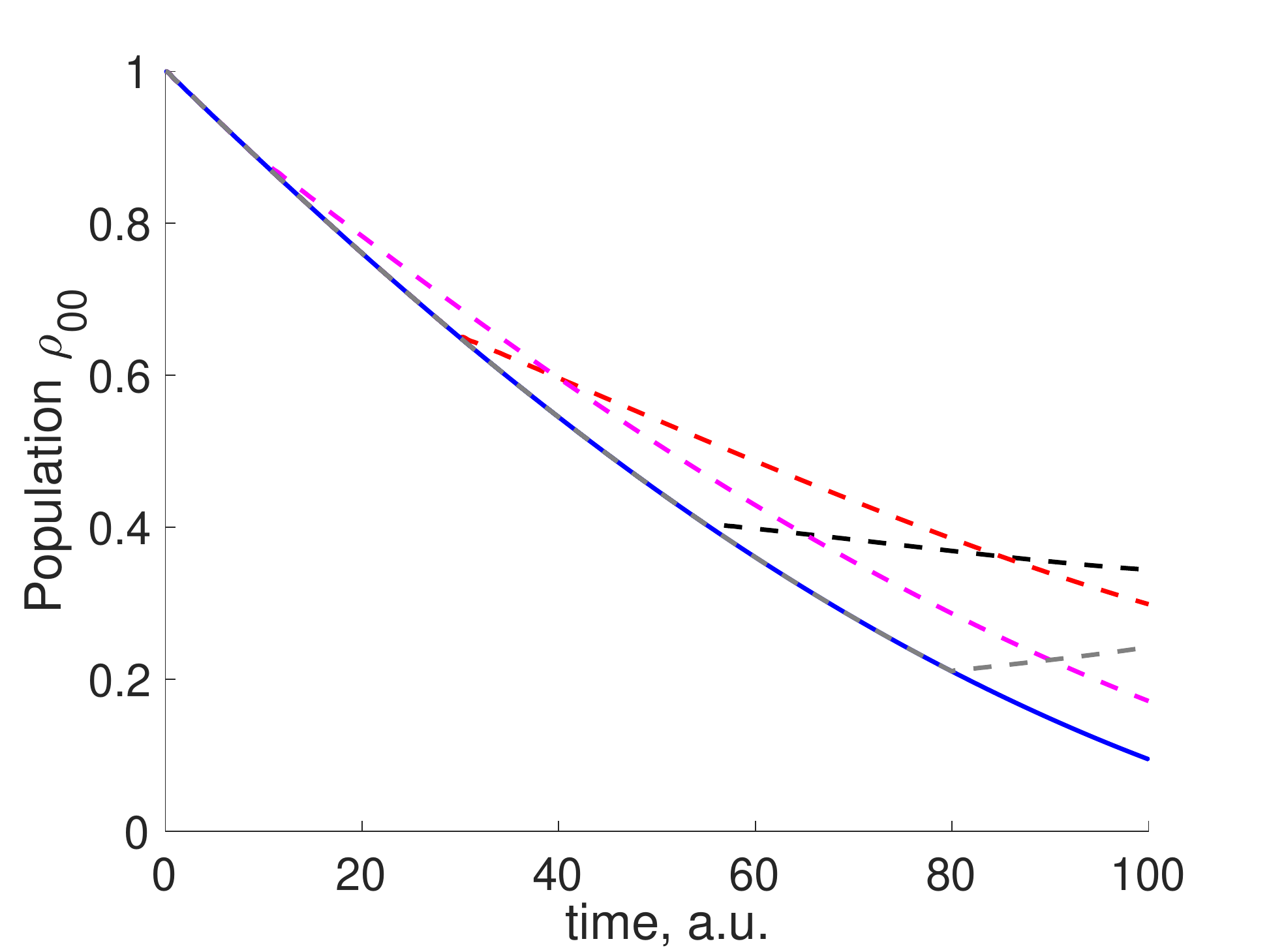}
    \caption{Population $\rho_{00}$ given by exact dynamics (blue solid) and four $U$ induced coherence 
    sign-flip dynamics that are different in the time of $U$ application: 10 a.u. (magenta dashed), 30 a.u. (red dashed), 55 a.u. (black dashed), and 80 a.u. (grey dashed).}
    \label{fig:LCUZeno}
\end{figure}  
\begin{figure}[h!]
    \centering
    \includegraphics[width=1.0\linewidth]{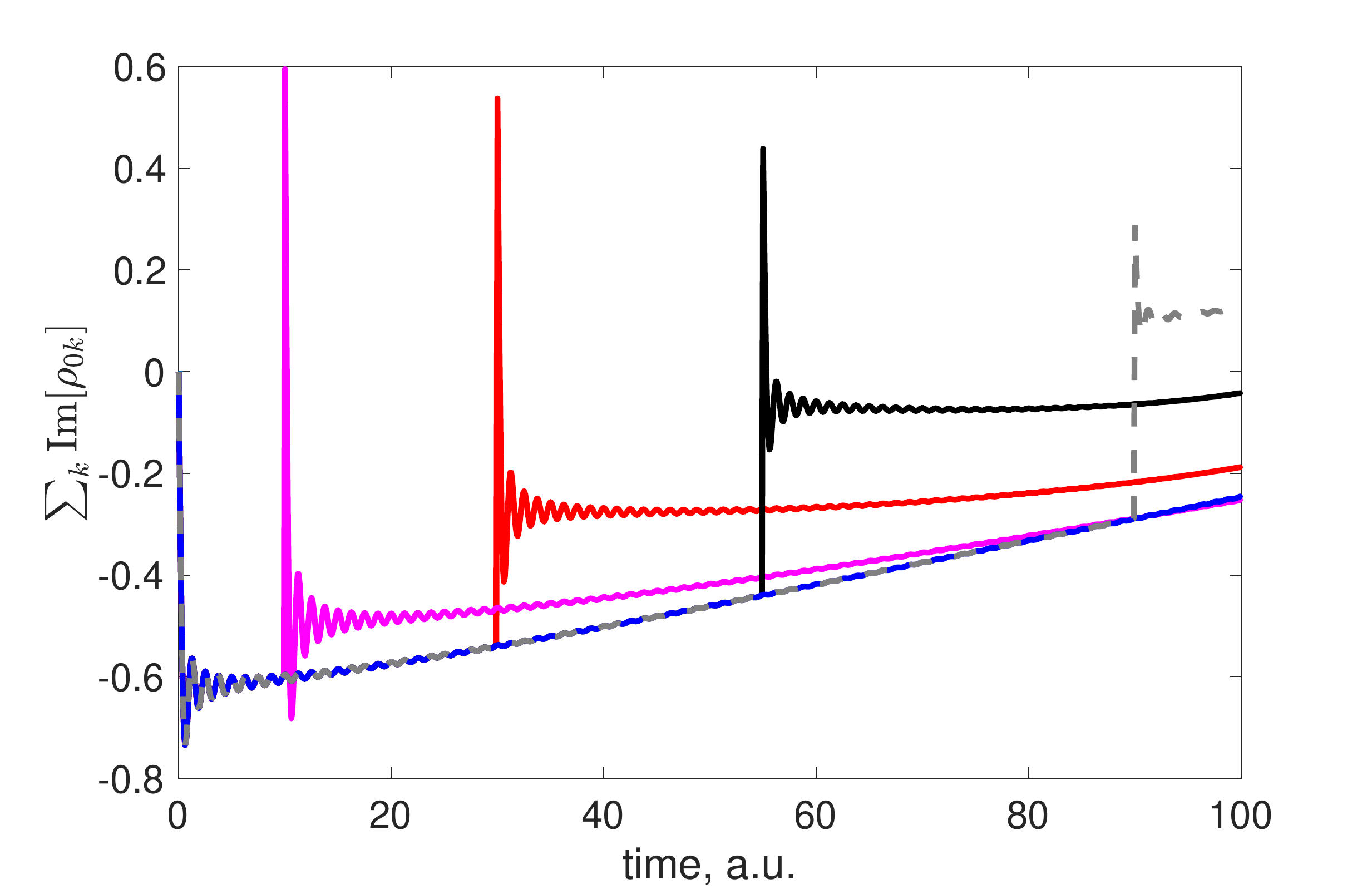}
    \caption{Sum of imaginary part of coherences between the level and continuum states, given by exact dynamics (blue solid) and four $U$ induced coherence 
    sign-flip dynamics that are different in the time of $U$ application: 10 a.u. (magenta dashed), 30 a.u. (red dashed), 55 a.u. (black dashed), and 90 a.u. (grey dashed).}
    \label{fig:LICCUZ}
\end{figure}  
For the experiment with the unitary transformation that changes the signs of all coherences between the 
level and the continuum, $U = 1-2\ket{0}\bra{0}$, trends in population (\fig{fig:LCUZeno}) 
and coherence (\fig{fig:LICCUZ}) dynamics are similar to those where measurements are used 
for disrupting the dynamics (\fig{fig:LCZeno} and \fig{fig:LICCZ}). 
The main difference is somewhat stronger QZE that can be connected to higher
levels of $\Sigma(t)$ minima right after altering the sign of coherences. This can be attributed to higher 
initial values of imaginary parts of coherences right after the $U$ action and 
the second term in \eq{eq:ImCtd}, this term was zero when the coherences vanished due to measurements. 
%${\rm Im}[\rho_{0k}]$ dynamics,    

\begin{figure}[h!]
    \centering
    \includegraphics[width=1.0\linewidth]{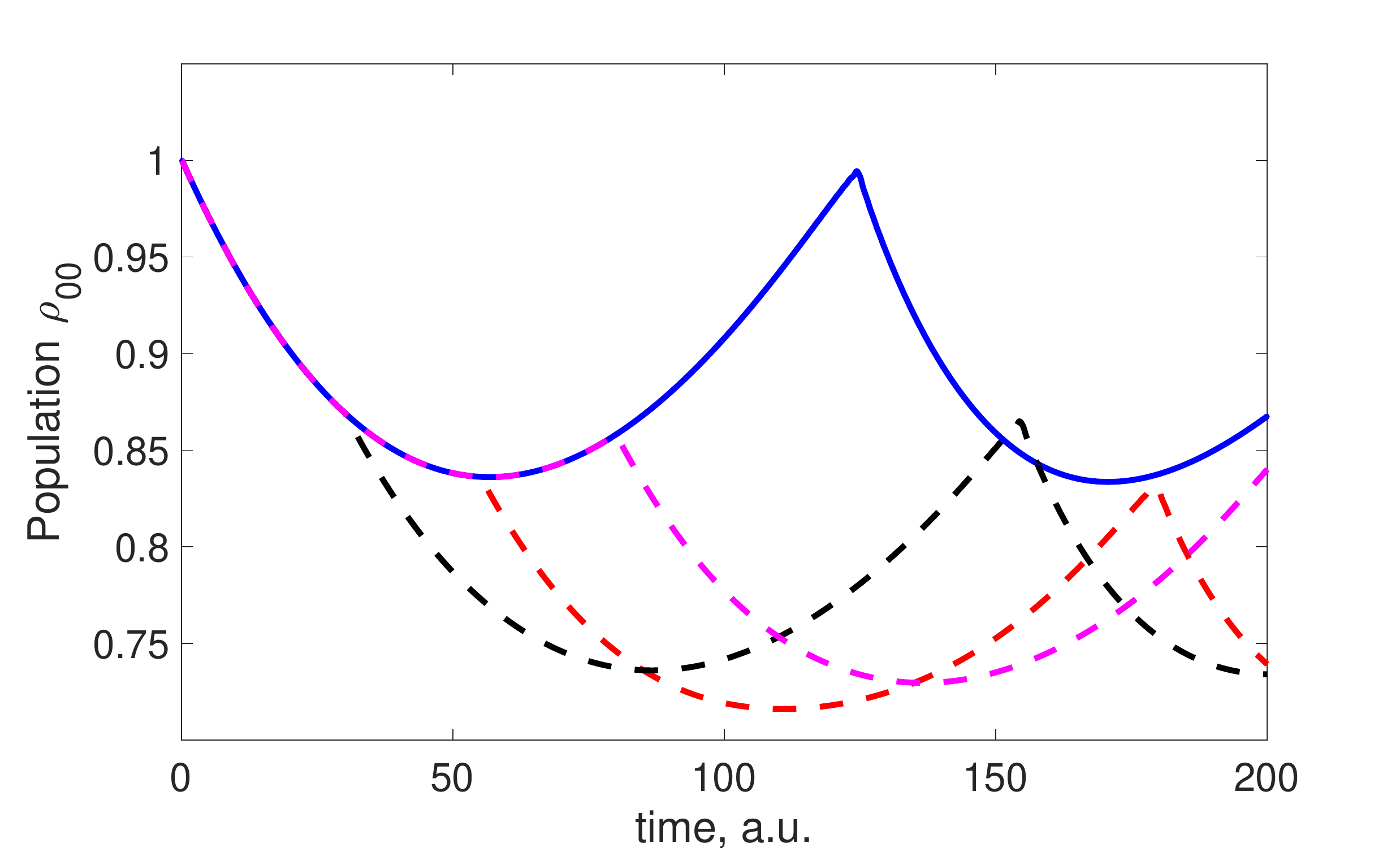}
    \caption{Population $\rho_{00}$ given by exact dynamics (blue solid) and four measurement 
    induced decoherences that are done in different times: 30 a.u. (black dashed), 55 a.u. (red dashed),  and 80 a.u. (magenta dashed).  }
    \label{fig:LoCPAZ}
\end{figure}

\begin{figure}[h!]
    \centering
    \includegraphics[width=1.0\linewidth]{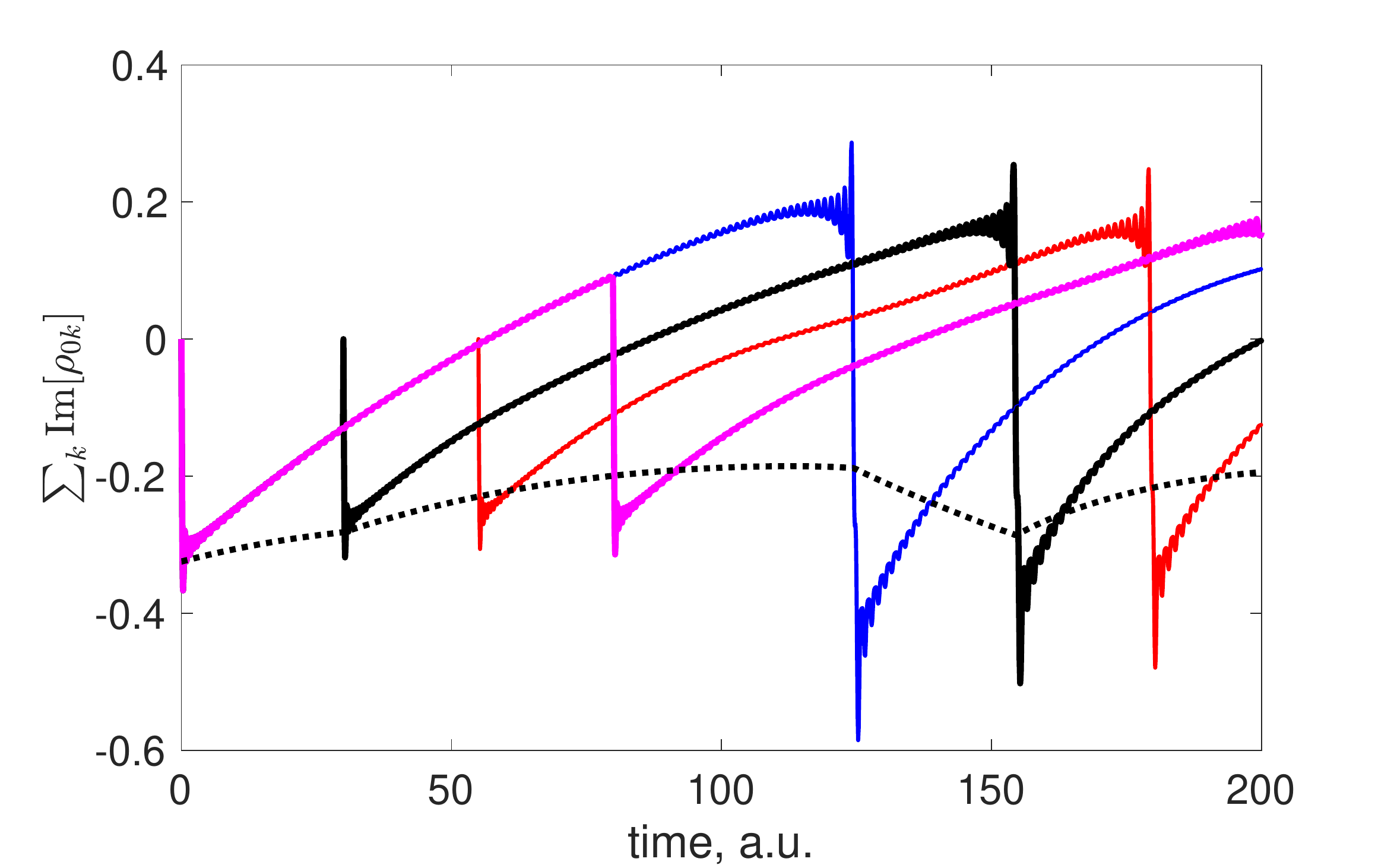}
    \caption{Sum of imaginary part of coherences between the level and continuum states, given by exact dynamics (blue solid) and four measurement induced decoherences that are done in different times: 30 a.u. (black solid), 55 a.u. (red solid),  and 80 a.u. (magenta solid).  The dotted black line is 
    $\Sigma^{(1)}(t_{\min})$ from \eq{eq:St_m}, its time dependence comes from $\DR{k}$ evaluated along the exact dynamics.}
    \label{fig:LoCCAZ}
\end{figure}

\paragraph{The level outside of a continuum model:} Since $\ket{0}$ level's energy $\epsilon_0$ 
is further away energetically from energies of levels representing the continuum, in this case 
the population dynamics oscillates between 1 and 0.83 in $\rho_{00}$ (\fig{fig:LoCPAZ}). 
When measurements applied 
along the dynamics they lead to the AZE. What are the reasons for this switch?
Considering $\Sigma(t)$ dynamics helps to provide the answer to this question (\fig{fig:LoCCAZ}).   
The main difference between the LOC and LIC models is that in LOC $\Sigma^{(1)}(t_{\min})$ stays at 
approximately the same level for all times and this level is below the rise of $\Sigma(t)$ due to 
natural dynamical decoherence described by long term limit of \eq{eq:SigmaT} 
(compare blue solid and dotted black lines on \fig{fig:LoCCAZ}). 
Thus, every time we measure $\rho_{00}$ in LOC, $\Sigma(t)$ very quickly goes to 
$\approx\Sigma^{(1)}(t_{\min})$, which stimulates the coherence channel of the population 
transfer. This relation is the opposite to the one we observed in the LIC model,
there $\Sigma^{(1)}(t_{\min})>\Sigma(t)$ (compare blue solid and dotted black lines on \fig{fig:LICCZ}).    

\begin{figure}[h!]
    \centering
    \includegraphics[width=1.0\linewidth]{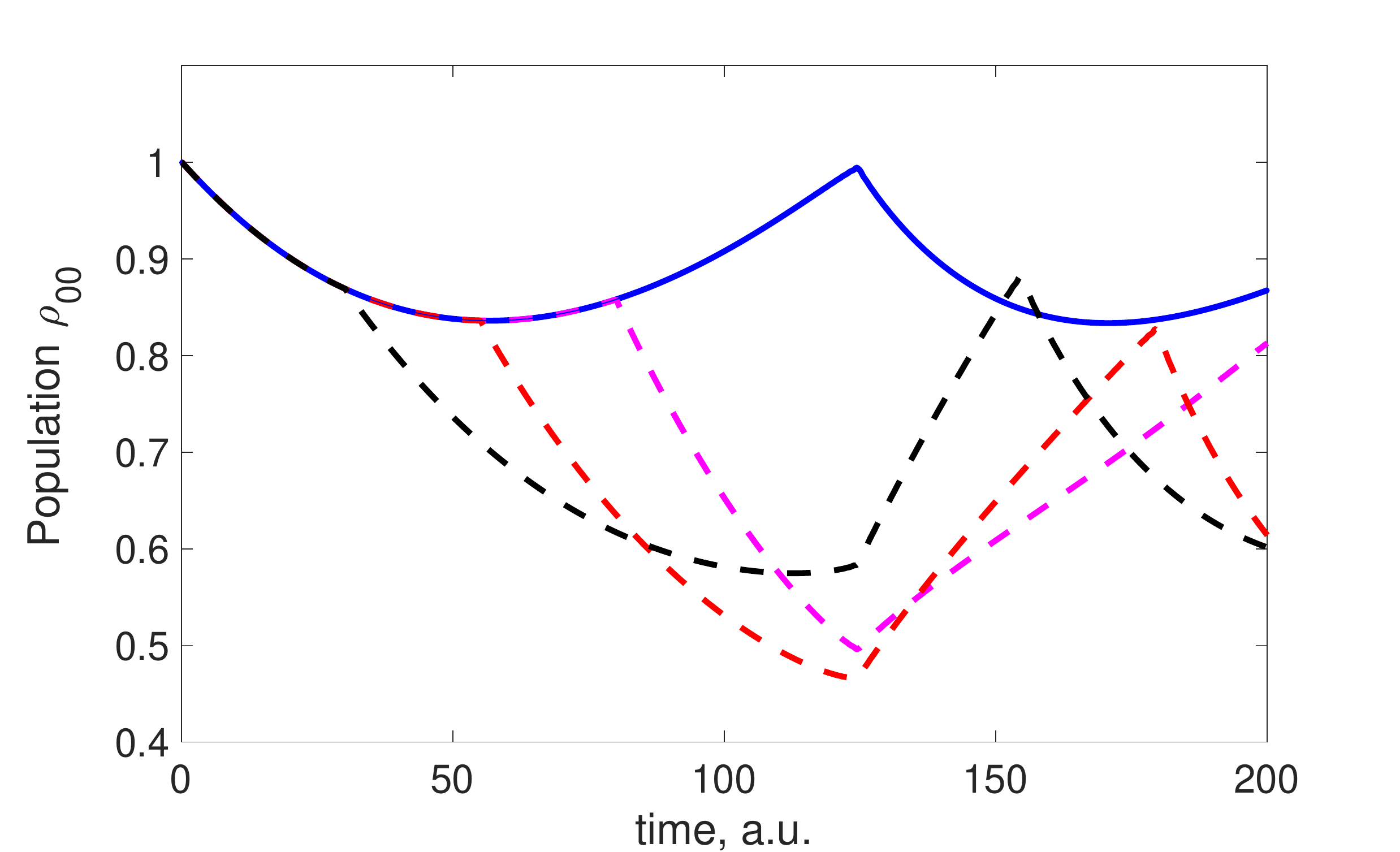}
    \caption{Population $\rho_{00}$ given by exact dynamics (blue solid) and four $U$ induced coherence 
    sign-flip dynamics that are different in the time of $U$ application: 30 a.u. (black dashed), 55 a.u. (red dashed),  and 80 a.u. (magenta dashed).}
    \label{fig:LoCPUAZ}
\end{figure}

\begin{figure}[h!]
    \centering
    \includegraphics[width=1.0\linewidth]{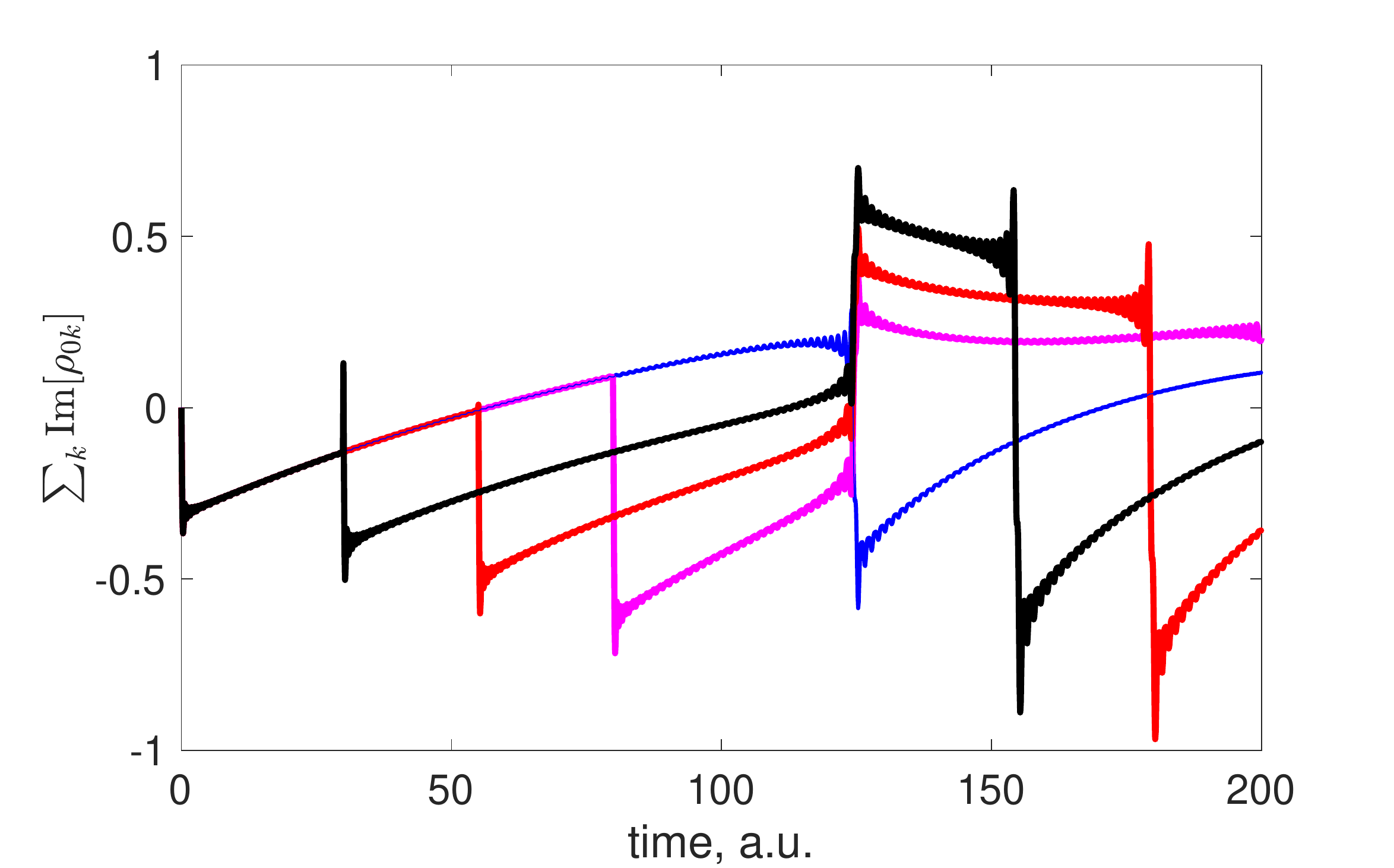}
    \caption{Sum of imaginary part of coherences between the level and continuum states, given by exact dynamics (blue solid) and four $U$ induced coherence 
    sign-flip dynamics that are different in the time of $U$ application: 30 a.u. (black solid), 55 a.u. (red solid),  and 80 a.u. (magenta solid).}
    \label{fig:LoCCUZ}
\end{figure}

When coherent control of LOC dynamics is done by $U$, the trends in population 
(\fig{fig:LoCPUAZ}) and coherence (\fig{fig:LoCCUZ})
dynamics are more intense compared to those when measurements were used: 
reduction of $\Sigma$ after the sign flip is lower than after zeroing it with the measurement, 
so the AZE is even more pronounced. We attribute this to similar factors as in the LIC 
model, not zeroing the coherences but rather changing their signs changes the initial conditions 
for dynamics after the $U$ influence and opens another channel for $\Sigma$ modification 
(the second term in \eq{eq:ImCtd}).

\section{Conclusions}

In this work we have shown on a microscopic level using simple models how the quantum 
Zeno and anti-Zeno effects operate. 
For the first time, a simple picture involving density matrix perturbation theory in low orders is used to 
explain coherent control that leads to these effects. The two scenarios that were considered are either 
zeroing coherences by doing measurements or altering signs of coherences by performing a unitary transformation. Qualitatively, these two control scenarios gave the same results, the main difference
was only quantitative, the unitary transformation usually intensifies either 
quantum-Zeno or anti-Zeno effect that 
can be induced by the measurement decoherence. The situation is more interesting in consideration 
of whether one would observe the 
quantum-Zeno or anti-Zeno effect in a particular setup. We have reached the 
conclusion that using only perturbative considerations and accurate values of state populations are 
sufficient to determine what effect will be observed.  If one needs to predict this for longer time-scales it is
safer to take state populations from the exact dynamical simulations.  

Both quantum-Zeno and anti-Zeno effects are examples of the coherent control because they originate 
from changes in quantum coherences that are induced by the measurement or application of the sign-changing 
unitary transformation. To understand both effects, one simply needs to realize that in the perturbation theory picture there are two channels of the transfer for population of one level to population of other levels:
1) the direct second-order transfer of population to population and 
2) the two-step transfer, where each steps 
are of the first order in perturbation, first, from population to coherence and then from coherence to population.
Coherent control processes discussed in this work alter the second channel, when this alternation leads 
to reduction of the coherence channel capacity 
this gives rise to the quantum-Zeno effect, otherwise we have the anti-Zeno effect.
Interestingly, the same manipulations done at different times or on different systems can lead to the 
opposite effects. This can be understood simply because coherences between any two states can 
stimulate both forward and backward transfers, depending on the sign of their imaginary part. 
Also, in general, coherences undergo their own dynamics, therefore they provide a flexible framework 
for controlling population transfer processes.\\    

\section*{Acknowledgements}
A.F.I. is grateful to Jeremy Schofield for stimulating conversations and 
thanks the Natural Sciences and Engineering Research Council of Canada for financial support.

%\bibliography{sn-bibliography}

\begin{thebibliography}{71}%
\makeatletter
\providecommand \@ifxundefined [1]{%
 \@ifx{#1\undefined}
}%
\providecommand \@ifnum [1]{%
 \ifnum #1\expandafter \@firstoftwo
 \else \expandafter \@secondoftwo
 \fi
}%
\providecommand \@ifx [1]{%
 \ifx #1\expandafter \@firstoftwo
 \else \expandafter \@secondoftwo
 \fi
}%
\providecommand \natexlab [1]{#1}%
\providecommand \enquote  [1]{``#1''}%
\providecommand \bibnamefont  [1]{#1}%
\providecommand \bibfnamefont [1]{#1}%
\providecommand \citenamefont [1]{#1}%
\providecommand \href@noop [0]{\@secondoftwo}%
\providecommand \href [0]{\begingroup \@sanitize@url \@href}%
\providecommand \@href[1]{\@@startlink{#1}\@@href}%
\providecommand \@@href[1]{\endgroup#1\@@endlink}%
\providecommand \@sanitize@url [0]{\catcode `\\12\catcode `\$12\catcode
  `\&12\catcode `\#12\catcode `\^12\catcode `\_12\catcode `\%12\relax}%
\providecommand \@@startlink[1]{}%
\providecommand \@@endlink[0]{}%
\providecommand \url  [0]{\begingroup\@sanitize@url \@url }%
\providecommand \@url [1]{\endgroup\@href {#1}{\urlprefix }}%
\providecommand \urlprefix  [0]{URL }%
\providecommand \Eprint [0]{\href }%
\providecommand \doibase [0]{http://dx.doi.org/}%
\providecommand \selectlanguage [0]{\@gobble}%
\providecommand \bibinfo  [0]{\@secondoftwo}%
\providecommand \bibfield  [0]{\@secondoftwo}%
\providecommand \translation [1]{[#1]}%
\providecommand \BibitemOpen [0]{}%
\providecommand \bibitemStop [0]{}%
\providecommand \bibitemNoStop [0]{.\EOS\space}%
\providecommand \EOS [0]{\spacefactor3000\relax}%
\providecommand \BibitemShut  [1]{\csname bibitem#1\endcsname}%
\let\auto@bib@innerbib\@empty
%</preamble>
\bibitem [{\citenamefont {Milburn}(1997)}]{Milburn1997-yb}%
  \BibitemOpen
  \bibfield  {author} {\bibinfo {author} {\bibfnamefont {G.~J.}\ \bibnamefont
  {Milburn}},\ }\href@noop {} {\emph {\bibinfo {title} {Schr{\"o}dinger's
  Machines: Machines: The Quantum Technology Reshaping Everyday Life}}}\
  (\bibinfo  {publisher} {W.H. Freeman},\ \bibinfo {address} {New York, NY},\
  \bibinfo {year} {1997})\BibitemShut {NoStop}%
\bibitem [{\citenamefont {Misra}\ and\ \citenamefont
  {Sudarshan}(1977)}]{misra1977zeno}%
  \BibitemOpen
  \bibfield  {author} {\bibinfo {author} {\bibfnamefont {B.}~\bibnamefont
  {Misra}}\ and\ \bibinfo {author} {\bibfnamefont {E.~G.}\ \bibnamefont
  {Sudarshan}},\ }\href@noop {} {\bibfield  {journal} {\bibinfo  {journal}
  {Journal of Mathematical Physics}\ }\textbf {\bibinfo {volume} {18}},\
  \bibinfo {pages} {756} (\bibinfo {year} {1977})}\BibitemShut {NoStop}%
\bibitem [{\citenamefont {Itano}\ \emph {et~al.}(1990)\citenamefont {Itano},
  \citenamefont {Heinzen}, \citenamefont {Bollinger},\ and\ \citenamefont
  {Wineland}}]{itano1990quantum}%
  \BibitemOpen
  \bibfield  {author} {\bibinfo {author} {\bibfnamefont {W.~M.}\ \bibnamefont
  {Itano}}, \bibinfo {author} {\bibfnamefont {D.~J.}\ \bibnamefont {Heinzen}},
  \bibinfo {author} {\bibfnamefont {J.~J.}\ \bibnamefont {Bollinger}}, \ and\
  \bibinfo {author} {\bibfnamefont {D.~J.}\ \bibnamefont {Wineland}},\
  }\href@noop {} {\bibfield  {journal} {\bibinfo  {journal} {Physical Review
  A}\ }\textbf {\bibinfo {volume} {41}},\ \bibinfo {pages} {2295} (\bibinfo
  {year} {1990})}\BibitemShut {NoStop}%
\bibitem [{\citenamefont {Kofman}\ and\ \citenamefont
  {Kurizki}(1996)}]{kofman1996quantum}%
  \BibitemOpen
  \bibfield  {author} {\bibinfo {author} {\bibfnamefont {A.}~\bibnamefont
  {Kofman}}\ and\ \bibinfo {author} {\bibfnamefont {G.}~\bibnamefont
  {Kurizki}},\ }\href@noop {} {\bibfield  {journal} {\bibinfo  {journal}
  {Physical Review A}\ }\textbf {\bibinfo {volume} {54}},\ \bibinfo {pages}
  {R3750} (\bibinfo {year} {1996})}\BibitemShut {NoStop}%
\bibitem [{\citenamefont {Home}\ and\ \citenamefont
  {Whitaker}(1997)}]{home1997conceptual}%
  \BibitemOpen
  \bibfield  {author} {\bibinfo {author} {\bibfnamefont {D.}~\bibnamefont
  {Home}}\ and\ \bibinfo {author} {\bibfnamefont {M.}~\bibnamefont
  {Whitaker}},\ }\href@noop {} {\bibfield  {journal} {\bibinfo  {journal}
  {Annals of Physics}\ }\textbf {\bibinfo {volume} {258}},\ \bibinfo {pages}
  {237} (\bibinfo {year} {1997})}\BibitemShut {NoStop}%
\bibitem [{\citenamefont {Koshino}\ and\ \citenamefont
  {Shimizu}(2005)}]{koshino2005quantum}%
  \BibitemOpen
  \bibfield  {author} {\bibinfo {author} {\bibfnamefont {K.}~\bibnamefont
  {Koshino}}\ and\ \bibinfo {author} {\bibfnamefont {A.}~\bibnamefont
  {Shimizu}},\ }\href@noop {} {\bibfield  {journal} {\bibinfo  {journal}
  {Physics reports}\ }\textbf {\bibinfo {volume} {412}},\ \bibinfo {pages}
  {191} (\bibinfo {year} {2005})}\BibitemShut {NoStop}%
\bibitem [{\citenamefont {Thapliyal}\ \emph {et~al.}(2016)\citenamefont
  {Thapliyal}, \citenamefont {Pathak},\ and\ \citenamefont
  {Pe{\v{r}}ina}}]{thapliyal2016linear}%
  \BibitemOpen
  \bibfield  {author} {\bibinfo {author} {\bibfnamefont {K.}~\bibnamefont
  {Thapliyal}}, \bibinfo {author} {\bibfnamefont {A.}~\bibnamefont {Pathak}}, \
  and\ \bibinfo {author} {\bibfnamefont {J.}~\bibnamefont {Pe{\v{r}}ina}},\
  }\href@noop {} {\bibfield  {journal} {\bibinfo  {journal} {Physical Review
  A}\ }\textbf {\bibinfo {volume} {93}},\ \bibinfo {pages} {022107} (\bibinfo
  {year} {2016})}\BibitemShut {NoStop}%
\bibitem [{\citenamefont {Gagen}\ and\ \citenamefont
  {Milburn}(1992)}]{gagen1992quantum}%
  \BibitemOpen
  \bibfield  {author} {\bibinfo {author} {\bibfnamefont {M.}~\bibnamefont
  {Gagen}}\ and\ \bibinfo {author} {\bibfnamefont {G.}~\bibnamefont
  {Milburn}},\ }\href@noop {} {\bibfield  {journal} {\bibinfo  {journal}
  {Physical Review A}\ }\textbf {\bibinfo {volume} {45}},\ \bibinfo {pages}
  {5228} (\bibinfo {year} {1992})}\BibitemShut {NoStop}%
\bibitem [{\citenamefont {Facchi}\ \emph {et~al.}(2004)\citenamefont {Facchi},
  \citenamefont {Lidar},\ and\ \citenamefont
  {Pascazio}}]{facchi2004unification}%
  \BibitemOpen
  \bibfield  {author} {\bibinfo {author} {\bibfnamefont {P.}~\bibnamefont
  {Facchi}}, \bibinfo {author} {\bibfnamefont {D.}~\bibnamefont {Lidar}}, \
  and\ \bibinfo {author} {\bibfnamefont {S.}~\bibnamefont {Pascazio}},\
  }\href@noop {} {\bibfield  {journal} {\bibinfo  {journal} {Physical Review
  A}\ }\textbf {\bibinfo {volume} {69}},\ \bibinfo {pages} {032314} (\bibinfo
  {year} {2004})}\BibitemShut {NoStop}%
\bibitem [{\citenamefont {Dhar}\ \emph {et~al.}(2005)\citenamefont {Dhar},
  \citenamefont {Grover},\ and\ \citenamefont {Roy}}]{dhar2005preserving}%
  \BibitemOpen
  \bibfield  {author} {\bibinfo {author} {\bibfnamefont {D.}~\bibnamefont
  {Dhar}}, \bibinfo {author} {\bibfnamefont {L.~K.}\ \bibnamefont {Grover}}, \
  and\ \bibinfo {author} {\bibfnamefont {S.~M.}\ \bibnamefont {Roy}},\
  }\href@noop {} {\bibfield  {journal} {\bibinfo  {journal} {arXiv preprint
  quant-ph/0504070}\ } (\bibinfo {year} {2005})}\BibitemShut {NoStop}%
\bibitem [{\citenamefont {Kofman}\ and\ \citenamefont
  {Kurizki}(2000)}]{kofman2000acceleration}%
  \BibitemOpen
  \bibfield  {author} {\bibinfo {author} {\bibfnamefont {A.}~\bibnamefont
  {Kofman}}\ and\ \bibinfo {author} {\bibfnamefont {G.}~\bibnamefont
  {Kurizki}},\ }\href@noop {} {\bibfield  {journal} {\bibinfo  {journal}
  {Nature}\ }\textbf {\bibinfo {volume} {405}},\ \bibinfo {pages} {546}
  (\bibinfo {year} {2000})}\BibitemShut {NoStop}%
\bibitem [{\citenamefont {Facchi}\ \emph {et~al.}(2001)\citenamefont {Facchi},
  \citenamefont {Nakazato},\ and\ \citenamefont
  {Pascazio}}]{facchi2001quantum}%
  \BibitemOpen
  \bibfield  {author} {\bibinfo {author} {\bibfnamefont {P.}~\bibnamefont
  {Facchi}}, \bibinfo {author} {\bibfnamefont {H.}~\bibnamefont {Nakazato}}, \
  and\ \bibinfo {author} {\bibfnamefont {S.}~\bibnamefont {Pascazio}},\
  }\href@noop {} {\bibfield  {journal} {\bibinfo  {journal} {Physical Review
  Letters}\ }\textbf {\bibinfo {volume} {86}},\ \bibinfo {pages} {2699}
  (\bibinfo {year} {2001})}\BibitemShut {NoStop}%
\bibitem [{\citenamefont {Yamaguchi}\ \emph {et~al.}(2008)\citenamefont
  {Yamaguchi}, \citenamefont {Asano},\ and\ \citenamefont
  {Noda}}]{yamaguchi2008photon}%
  \BibitemOpen
  \bibfield  {author} {\bibinfo {author} {\bibfnamefont {M.}~\bibnamefont
  {Yamaguchi}}, \bibinfo {author} {\bibfnamefont {T.}~\bibnamefont {Asano}}, \
  and\ \bibinfo {author} {\bibfnamefont {S.}~\bibnamefont {Noda}},\ }\href@noop
  {} {\bibfield  {journal} {\bibinfo  {journal} {Optics Express}\ }\textbf
  {\bibinfo {volume} {16}},\ \bibinfo {pages} {18067} (\bibinfo {year}
  {2008})}\BibitemShut {NoStop}%
\bibitem [{\citenamefont {Fischer}\ \emph {et~al.}(2001)\citenamefont
  {Fischer}, \citenamefont {Guti{\'e}rrez-Medina},\ and\ \citenamefont
  {Raizen}}]{fischer2001observation}%
  \BibitemOpen
  \bibfield  {author} {\bibinfo {author} {\bibfnamefont {M.~C.}\ \bibnamefont
  {Fischer}}, \bibinfo {author} {\bibfnamefont {B.}~\bibnamefont
  {Guti{\'e}rrez-Medina}}, \ and\ \bibinfo {author} {\bibfnamefont {M.~G.}\
  \bibnamefont {Raizen}},\ }\href@noop {} {\bibfield  {journal} {\bibinfo
  {journal} {Physical review letters}\ }\textbf {\bibinfo {volume} {87}},\
  \bibinfo {pages} {040402} (\bibinfo {year} {2001})}\BibitemShut {NoStop}%
\bibitem [{\citenamefont {Facchi}\ and\ \citenamefont
  {Pascazio}(2002)}]{facchi2002quantum}%
  \BibitemOpen
  \bibfield  {author} {\bibinfo {author} {\bibfnamefont {P.}~\bibnamefont
  {Facchi}}\ and\ \bibinfo {author} {\bibfnamefont {S.}~\bibnamefont
  {Pascazio}},\ }\href@noop {} {\bibfield  {journal} {\bibinfo  {journal}
  {Physical review letters}\ }\textbf {\bibinfo {volume} {89}},\ \bibinfo
  {pages} {080401} (\bibinfo {year} {2002})}\BibitemShut {NoStop}%
\bibitem [{\citenamefont {Sch{\"a}fer}\ \emph {et~al.}(2014)\citenamefont
  {Sch{\"a}fer}, \citenamefont {Herrera}, \citenamefont {Cherukattil},
  \citenamefont {Lovecchio}, \citenamefont {Cataliotti}, \citenamefont
  {Caruso},\ and\ \citenamefont {Smerzi}}]{schafer2014experimental}%
  \BibitemOpen
  \bibfield  {author} {\bibinfo {author} {\bibfnamefont {F.}~\bibnamefont
  {Sch{\"a}fer}}, \bibinfo {author} {\bibfnamefont {I.}~\bibnamefont
  {Herrera}}, \bibinfo {author} {\bibfnamefont {S.}~\bibnamefont
  {Cherukattil}}, \bibinfo {author} {\bibfnamefont {C.}~\bibnamefont
  {Lovecchio}}, \bibinfo {author} {\bibfnamefont {F.~S.}\ \bibnamefont
  {Cataliotti}}, \bibinfo {author} {\bibfnamefont {F.}~\bibnamefont {Caruso}},
  \ and\ \bibinfo {author} {\bibfnamefont {A.}~\bibnamefont {Smerzi}},\
  }\href@noop {} {\bibfield  {journal} {\bibinfo  {journal} {Nature
  communications}\ }\textbf {\bibinfo {volume} {5}},\ \bibinfo {pages} {3194}
  (\bibinfo {year} {2014})}\BibitemShut {NoStop}%
\bibitem [{\citenamefont {Signoles}\ \emph {et~al.}(2014)\citenamefont
  {Signoles}, \citenamefont {Facon}, \citenamefont {Grosso}, \citenamefont
  {Dotsenko}, \citenamefont {Haroche}, \citenamefont {Raimond}, \citenamefont
  {Brune},\ and\ \citenamefont {Gleyzes}}]{signoles2014confined}%
  \BibitemOpen
  \bibfield  {author} {\bibinfo {author} {\bibfnamefont {A.}~\bibnamefont
  {Signoles}}, \bibinfo {author} {\bibfnamefont {A.}~\bibnamefont {Facon}},
  \bibinfo {author} {\bibfnamefont {D.}~\bibnamefont {Grosso}}, \bibinfo
  {author} {\bibfnamefont {I.}~\bibnamefont {Dotsenko}}, \bibinfo {author}
  {\bibfnamefont {S.}~\bibnamefont {Haroche}}, \bibinfo {author} {\bibfnamefont
  {J.-M.}\ \bibnamefont {Raimond}}, \bibinfo {author} {\bibfnamefont
  {M.}~\bibnamefont {Brune}}, \ and\ \bibinfo {author} {\bibfnamefont
  {S.}~\bibnamefont {Gleyzes}},\ }\href@noop {} {\bibfield  {journal} {\bibinfo
   {journal} {Nature Physics}\ }\textbf {\bibinfo {volume} {10}},\ \bibinfo
  {pages} {715} (\bibinfo {year} {2014})}\BibitemShut {NoStop}%
\bibitem [{\citenamefont {Gherardini}\ \emph {et~al.}(2016)\citenamefont
  {Gherardini}, \citenamefont {Gupta}, \citenamefont {Cataliotti},
  \citenamefont {Smerzi}, \citenamefont {Caruso},\ and\ \citenamefont
  {Ruffo}}]{gherardini2016stochastic}%
  \BibitemOpen
  \bibfield  {author} {\bibinfo {author} {\bibfnamefont {S.}~\bibnamefont
  {Gherardini}}, \bibinfo {author} {\bibfnamefont {S.}~\bibnamefont {Gupta}},
  \bibinfo {author} {\bibfnamefont {F.~S.}\ \bibnamefont {Cataliotti}},
  \bibinfo {author} {\bibfnamefont {A.}~\bibnamefont {Smerzi}}, \bibinfo
  {author} {\bibfnamefont {F.}~\bibnamefont {Caruso}}, \ and\ \bibinfo {author}
  {\bibfnamefont {S.}~\bibnamefont {Ruffo}},\ }\href@noop {} {\bibfield
  {journal} {\bibinfo  {journal} {New Journal of Physics}\ }\textbf {\bibinfo
  {volume} {18}},\ \bibinfo {pages} {013048} (\bibinfo {year}
  {2016})}\BibitemShut {NoStop}%
\bibitem [{\citenamefont {M{\"u}ller}\ \emph {et~al.}(2016)\citenamefont
  {M{\"u}ller}, \citenamefont {Gherardini},\ and\ \citenamefont
  {Caruso}}]{muller2016stochastic}%
  \BibitemOpen
  \bibfield  {author} {\bibinfo {author} {\bibfnamefont {M.~M.}\ \bibnamefont
  {M{\"u}ller}}, \bibinfo {author} {\bibfnamefont {S.}~\bibnamefont
  {Gherardini}}, \ and\ \bibinfo {author} {\bibfnamefont {F.}~\bibnamefont
  {Caruso}},\ }\href@noop {} {\bibfield  {journal} {\bibinfo  {journal}
  {Scientific reports}\ }\textbf {\bibinfo {volume} {6}},\ \bibinfo {pages} {1}
  (\bibinfo {year} {2016})}\BibitemShut {NoStop}%
\bibitem [{\citenamefont {Facchi}\ and\ \citenamefont
  {Pascazio}(2008)}]{Facchi2008}%
  \BibitemOpen
  \bibfield  {author} {\bibinfo {author} {\bibfnamefont {P.}~\bibnamefont
  {Facchi}}\ and\ \bibinfo {author} {\bibfnamefont {S.}~\bibnamefont
  {Pascazio}},\ }\href {\doibase 10.1088/1751-8113/41/49/493001} {\bibfield
  {journal} {\bibinfo  {journal} {Journal of Physics A: Mathematical and
  Theoretical}\ }\textbf {\bibinfo {volume} {41}},\ \bibinfo {pages} {493001}
  (\bibinfo {year} {2008})}\BibitemShut {NoStop}%
\bibitem [{\citenamefont {Kwiat}\ \emph {et~al.}(1999)\citenamefont {Kwiat},
  \citenamefont {White}, \citenamefont {Mitchell}, \citenamefont {Nairz},
  \citenamefont {Weihs}, \citenamefont {Weinfurter},\ and\ \citenamefont
  {Zeilinger}}]{Kwiat1999}%
  \BibitemOpen
  \bibfield  {author} {\bibinfo {author} {\bibfnamefont {P.~G.}\ \bibnamefont
  {Kwiat}}, \bibinfo {author} {\bibfnamefont {A.~G.}\ \bibnamefont {White}},
  \bibinfo {author} {\bibfnamefont {J.~R.}\ \bibnamefont {Mitchell}}, \bibinfo
  {author} {\bibfnamefont {O.}~\bibnamefont {Nairz}}, \bibinfo {author}
  {\bibfnamefont {G.}~\bibnamefont {Weihs}}, \bibinfo {author} {\bibfnamefont
  {H.}~\bibnamefont {Weinfurter}}, \ and\ \bibinfo {author} {\bibfnamefont
  {A.}~\bibnamefont {Zeilinger}},\ }\href {\doibase
  10.1103/physrevlett.83.4725} {\bibfield  {journal} {\bibinfo  {journal}
  {Physical Review Letters}\ }\textbf {\bibinfo {volume} {83}},\ \bibinfo
  {pages} {4725} (\bibinfo {year} {1999})}\BibitemShut {NoStop}%
\bibitem [{\citenamefont {Zheng}\ \emph {et~al.}(2008)\citenamefont {Zheng},
  \citenamefont {Zhu},\ and\ \citenamefont {Zubairy}}]{Zheng2008}%
  \BibitemOpen
  \bibfield  {author} {\bibinfo {author} {\bibfnamefont {H.}~\bibnamefont
  {Zheng}}, \bibinfo {author} {\bibfnamefont {S.~Y.}\ \bibnamefont {Zhu}}, \
  and\ \bibinfo {author} {\bibfnamefont {M.~S.}\ \bibnamefont {Zubairy}},\
  }\href {\doibase 10.1103/physrevlett.101.200404} {\bibfield  {journal}
  {\bibinfo  {journal} {Physical Review Letters}\ }\textbf {\bibinfo {volume}
  {101}} (\bibinfo {year} {2008}),\ 10.1103/physrevlett.101.200404}\BibitemShut
  {NoStop}%
\bibitem [{\citenamefont {Virz{\`\i}}\ \emph {et~al.}(2022)\citenamefont
  {Virz{\`\i}}, \citenamefont {Avella}, \citenamefont {Piacentini},
  \citenamefont {Gramegna}, \citenamefont {Opatrn{\`y}}, \citenamefont
  {Kofman}, \citenamefont {Kurizki}, \citenamefont {Gherardini}, \citenamefont
  {Caruso}, \citenamefont {Degiovanni} \emph {et~al.}}]{virzi2022quantum}%
  \BibitemOpen
  \bibfield  {author} {\bibinfo {author} {\bibfnamefont {S.}~\bibnamefont
  {Virz{\`\i}}}, \bibinfo {author} {\bibfnamefont {A.}~\bibnamefont {Avella}},
  \bibinfo {author} {\bibfnamefont {F.}~\bibnamefont {Piacentini}}, \bibinfo
  {author} {\bibfnamefont {M.}~\bibnamefont {Gramegna}}, \bibinfo {author}
  {\bibfnamefont {T.}~\bibnamefont {Opatrn{\`y}}}, \bibinfo {author}
  {\bibfnamefont {A.~G.}\ \bibnamefont {Kofman}}, \bibinfo {author}
  {\bibfnamefont {G.}~\bibnamefont {Kurizki}}, \bibinfo {author} {\bibfnamefont
  {S.}~\bibnamefont {Gherardini}}, \bibinfo {author} {\bibfnamefont
  {F.}~\bibnamefont {Caruso}}, \bibinfo {author} {\bibfnamefont {I.~P.}\
  \bibnamefont {Degiovanni}},  \emph {et~al.},\ }\href@noop {} {\bibfield
  {journal} {\bibinfo  {journal} {Physical Review Letters}\ }\textbf {\bibinfo
  {volume} {129}},\ \bibinfo {pages} {030401} (\bibinfo {year}
  {2022})}\BibitemShut {NoStop}%
\bibitem [{\citenamefont {Do}\ \emph {et~al.}(2019)\citenamefont {Do},
  \citenamefont {Lovecchio}, \citenamefont {Mastroserio}, \citenamefont
  {Fabbri}, \citenamefont {Cataliotti}, \citenamefont {Gherardini},
  \citenamefont {M{\"u}ller}, \citenamefont {Dalla~Pozza},\ and\ \citenamefont
  {Caruso}}]{do2019experimental}%
  \BibitemOpen
  \bibfield  {author} {\bibinfo {author} {\bibfnamefont {H.-V.}\ \bibnamefont
  {Do}}, \bibinfo {author} {\bibfnamefont {C.}~\bibnamefont {Lovecchio}},
  \bibinfo {author} {\bibfnamefont {I.}~\bibnamefont {Mastroserio}}, \bibinfo
  {author} {\bibfnamefont {N.}~\bibnamefont {Fabbri}}, \bibinfo {author}
  {\bibfnamefont {F.~S.}\ \bibnamefont {Cataliotti}}, \bibinfo {author}
  {\bibfnamefont {S.}~\bibnamefont {Gherardini}}, \bibinfo {author}
  {\bibfnamefont {M.~M.}\ \bibnamefont {M{\"u}ller}}, \bibinfo {author}
  {\bibfnamefont {N.}~\bibnamefont {Dalla~Pozza}}, \ and\ \bibinfo {author}
  {\bibfnamefont {F.}~\bibnamefont {Caruso}},\ }\href@noop {} {\bibfield
  {journal} {\bibinfo  {journal} {New Journal of Physics}\ }\textbf {\bibinfo
  {volume} {21}},\ \bibinfo {pages} {113056} (\bibinfo {year}
  {2019})}\BibitemShut {NoStop}%
\bibitem [{\citenamefont {Long}\ \emph {et~al.}(2022)\citenamefont {Long},
  \citenamefont {He}, \citenamefont {Zhang}, \citenamefont {Tang},
  \citenamefont {Lin}, \citenamefont {Liu}, \citenamefont {Nie}, \citenamefont
  {Feng}, \citenamefont {Li}, \citenamefont {Xin} \emph
  {et~al.}}]{long2022entanglement}%
  \BibitemOpen
  \bibfield  {author} {\bibinfo {author} {\bibfnamefont {X.}~\bibnamefont
  {Long}}, \bibinfo {author} {\bibfnamefont {W.-T.}\ \bibnamefont {He}},
  \bibinfo {author} {\bibfnamefont {N.-N.}\ \bibnamefont {Zhang}}, \bibinfo
  {author} {\bibfnamefont {K.}~\bibnamefont {Tang}}, \bibinfo {author}
  {\bibfnamefont {Z.}~\bibnamefont {Lin}}, \bibinfo {author} {\bibfnamefont
  {H.}~\bibnamefont {Liu}}, \bibinfo {author} {\bibfnamefont {X.}~\bibnamefont
  {Nie}}, \bibinfo {author} {\bibfnamefont {G.}~\bibnamefont {Feng}}, \bibinfo
  {author} {\bibfnamefont {J.}~\bibnamefont {Li}}, \bibinfo {author}
  {\bibfnamefont {T.}~\bibnamefont {Xin}},  \emph {et~al.},\ }\href@noop {}
  {\bibfield  {journal} {\bibinfo  {journal} {Physical Review Letters}\
  }\textbf {\bibinfo {volume} {129}},\ \bibinfo {pages} {070502} (\bibinfo
  {year} {2022})}\BibitemShut {NoStop}%
\bibitem [{\citenamefont {Wu}\ \emph {et~al.}(2020)\citenamefont {Wu},
  \citenamefont {Wang}, \citenamefont {Su}, \citenamefont {Xia}, \citenamefont
  {Jiang},\ and\ \citenamefont {Song}}]{wu2020discrimination}%
  \BibitemOpen
  \bibfield  {author} {\bibinfo {author} {\bibfnamefont {J.-L.}\ \bibnamefont
  {Wu}}, \bibinfo {author} {\bibfnamefont {Y.}~\bibnamefont {Wang}}, \bibinfo
  {author} {\bibfnamefont {S.-L.}\ \bibnamefont {Su}}, \bibinfo {author}
  {\bibfnamefont {Y.}~\bibnamefont {Xia}}, \bibinfo {author} {\bibfnamefont
  {Y.}~\bibnamefont {Jiang}}, \ and\ \bibinfo {author} {\bibfnamefont
  {J.}~\bibnamefont {Song}},\ }\href@noop {} {\bibfield  {journal} {\bibinfo
  {journal} {Optics Express}\ }\textbf {\bibinfo {volume} {28}},\ \bibinfo
  {pages} {33475} (\bibinfo {year} {2020})}\BibitemShut {NoStop}%
\bibitem [{\citenamefont {Liu}\ \emph {et~al.}(2023)\citenamefont {Liu},
  \citenamefont {Liu}, \citenamefont {Ziegler},\ and\ \citenamefont
  {Chen}}]{liu2023engineering}%
  \BibitemOpen
  \bibfield  {author} {\bibinfo {author} {\bibfnamefont {Q.}~\bibnamefont
  {Liu}}, \bibinfo {author} {\bibfnamefont {W.}~\bibnamefont {Liu}}, \bibinfo
  {author} {\bibfnamefont {K.}~\bibnamefont {Ziegler}}, \ and\ \bibinfo
  {author} {\bibfnamefont {F.}~\bibnamefont {Chen}},\ }\href@noop {} {\bibfield
   {journal} {\bibinfo  {journal} {Physical Review Letters}\ }\textbf {\bibinfo
  {volume} {130}},\ \bibinfo {pages} {103801} (\bibinfo {year}
  {2023})}\BibitemShut {NoStop}%
\bibitem [{\citenamefont {Longhi}(2006)}]{longhi2006nonexponential}%
  \BibitemOpen
  \bibfield  {author} {\bibinfo {author} {\bibfnamefont {S.}~\bibnamefont
  {Longhi}},\ }\href@noop {} {\bibfield  {journal} {\bibinfo  {journal}
  {Physical review letters}\ }\textbf {\bibinfo {volume} {97}},\ \bibinfo
  {pages} {110402} (\bibinfo {year} {2006})}\BibitemShut {NoStop}%
\bibitem [{\citenamefont {Das}\ \emph {et~al.}(2021)\citenamefont {Das},
  \citenamefont {Thapliyal}, \citenamefont {Sen}, \citenamefont
  {Pe{\v{r}}ina},\ and\ \citenamefont {Pathak}}]{das2021interplay}%
  \BibitemOpen
  \bibfield  {author} {\bibinfo {author} {\bibfnamefont {M.}~\bibnamefont
  {Das}}, \bibinfo {author} {\bibfnamefont {K.}~\bibnamefont {Thapliyal}},
  \bibinfo {author} {\bibfnamefont {B.}~\bibnamefont {Sen}}, \bibinfo {author}
  {\bibfnamefont {J.}~\bibnamefont {Pe{\v{r}}ina}}, \ and\ \bibinfo {author}
  {\bibfnamefont {A.}~\bibnamefont {Pathak}},\ }\href@noop {} {\bibfield
  {journal} {\bibinfo  {journal} {Physical Review A}\ }\textbf {\bibinfo
  {volume} {103}},\ \bibinfo {pages} {013713} (\bibinfo {year}
  {2021})}\BibitemShut {NoStop}%
\bibitem [{\citenamefont {Das}\ \emph {et~al.}(2023)\citenamefont {Das},
  \citenamefont {Sen}, \citenamefont {Thapliyal},\ and\ \citenamefont
  {Pathak}}]{das2023quantum}%
  \BibitemOpen
  \bibfield  {author} {\bibinfo {author} {\bibfnamefont {M.}~\bibnamefont
  {Das}}, \bibinfo {author} {\bibfnamefont {B.}~\bibnamefont {Sen}}, \bibinfo
  {author} {\bibfnamefont {K.}~\bibnamefont {Thapliyal}}, \ and\ \bibinfo
  {author} {\bibfnamefont {A.}~\bibnamefont {Pathak}},\ }\href@noop {}
  {\bibfield  {journal} {\bibinfo  {journal} {arXiv preprint arXiv:2304.04073}\
  } (\bibinfo {year} {2023})}\BibitemShut {NoStop}%
\bibitem [{\citenamefont {Zezyulin}\ \emph {et~al.}(2012)\citenamefont
  {Zezyulin}, \citenamefont {Konotop}, \citenamefont {Barontini},\ and\
  \citenamefont {Ott}}]{zezyulin2012macroscopic}%
  \BibitemOpen
  \bibfield  {author} {\bibinfo {author} {\bibfnamefont {D.}~\bibnamefont
  {Zezyulin}}, \bibinfo {author} {\bibfnamefont {V.}~\bibnamefont {Konotop}},
  \bibinfo {author} {\bibfnamefont {G.}~\bibnamefont {Barontini}}, \ and\
  \bibinfo {author} {\bibfnamefont {H.}~\bibnamefont {Ott}},\ }\href@noop {}
  {\bibfield  {journal} {\bibinfo  {journal} {Physical review letters}\
  }\textbf {\bibinfo {volume} {109}},\ \bibinfo {pages} {020405} (\bibinfo
  {year} {2012})}\BibitemShut {NoStop}%
\bibitem [{\citenamefont {Huang}\ \emph {et~al.}(2010)\citenamefont {Huang},
  \citenamefont {Altepeter},\ and\ \citenamefont
  {Kumar}}]{huang2010interaction}%
  \BibitemOpen
  \bibfield  {author} {\bibinfo {author} {\bibfnamefont {Y.-P.}\ \bibnamefont
  {Huang}}, \bibinfo {author} {\bibfnamefont {J.~B.}\ \bibnamefont
  {Altepeter}}, \ and\ \bibinfo {author} {\bibfnamefont {P.}~\bibnamefont
  {Kumar}},\ }\href@noop {} {\bibfield  {journal} {\bibinfo  {journal}
  {Physical Review A}\ }\textbf {\bibinfo {volume} {82}},\ \bibinfo {pages}
  {063826} (\bibinfo {year} {2010})}\BibitemShut {NoStop}%
\bibitem [{\citenamefont {Nodurft}\ \emph
  {et~al.}(2022{\natexlab{a}})\citenamefont {Nodurft}, \citenamefont {Kirby},
  \citenamefont {Glasser}, \citenamefont {Shaw},\ and\ \citenamefont
  {Searles}}]{nodurft2022polarization}%
  \BibitemOpen
  \bibfield  {author} {\bibinfo {author} {\bibfnamefont {I.}~\bibnamefont
  {Nodurft}}, \bibinfo {author} {\bibfnamefont {B.}~\bibnamefont {Kirby}},
  \bibinfo {author} {\bibfnamefont {R.}~\bibnamefont {Glasser}}, \bibinfo
  {author} {\bibfnamefont {H.}~\bibnamefont {Shaw}}, \ and\ \bibinfo {author}
  {\bibfnamefont {T.}~\bibnamefont {Searles}},\ }in\ \href@noop {} {\emph
  {\bibinfo {booktitle} {Quantum 2.0}}}\ (\bibinfo {organization} {Optica
  Publishing Group},\ \bibinfo {year} {2022})\ pp.\ \bibinfo {pages}
  {QTu2A--5}\BibitemShut {NoStop}%
\bibitem [{\citenamefont {Nodurft}\ \emph
  {et~al.}(2022{\natexlab{b}})\citenamefont {Nodurft}, \citenamefont {Shaw},
  \citenamefont {Glasser}, \citenamefont {Kirby},\ and\ \citenamefont
  {Searles}}]{nodurft2022generation}%
  \BibitemOpen
  \bibfield  {author} {\bibinfo {author} {\bibfnamefont {I.~C.}\ \bibnamefont
  {Nodurft}}, \bibinfo {author} {\bibfnamefont {H.~C.}\ \bibnamefont {Shaw}},
  \bibinfo {author} {\bibfnamefont {R.~T.}\ \bibnamefont {Glasser}}, \bibinfo
  {author} {\bibfnamefont {B.~T.}\ \bibnamefont {Kirby}}, \ and\ \bibinfo
  {author} {\bibfnamefont {T.~A.}\ \bibnamefont {Searles}},\ }\href@noop {}
  {\bibfield  {journal} {\bibinfo  {journal} {Optics Express}\ }\textbf
  {\bibinfo {volume} {30}},\ \bibinfo {pages} {31971} (\bibinfo {year}
  {2022}{\natexlab{b}})}\BibitemShut {NoStop}%
\bibitem [{\citenamefont {Jacobs}\ and\ \citenamefont
  {Franson}(2009)}]{jacobs2009all}%
  \BibitemOpen
  \bibfield  {author} {\bibinfo {author} {\bibfnamefont {B.~C.}\ \bibnamefont
  {Jacobs}}\ and\ \bibinfo {author} {\bibfnamefont {J.}~\bibnamefont
  {Franson}},\ }\href@noop {} {\bibfield  {journal} {\bibinfo  {journal}
  {Physical Review A}\ }\textbf {\bibinfo {volume} {79}},\ \bibinfo {pages}
  {063830} (\bibinfo {year} {2009})}\BibitemShut {NoStop}%
\bibitem [{\citenamefont {Zhou}\ \emph {et~al.}(2009)\citenamefont {Zhou},
  \citenamefont {Yang}, \citenamefont {Liu}, \citenamefont {Sun}, \citenamefont
  {Nori} \emph {et~al.}}]{zhou2009quantum}%
  \BibitemOpen
  \bibfield  {author} {\bibinfo {author} {\bibfnamefont {L.}~\bibnamefont
  {Zhou}}, \bibinfo {author} {\bibfnamefont {S.}~\bibnamefont {Yang}}, \bibinfo
  {author} {\bibfnamefont {Y.-x.}\ \bibnamefont {Liu}}, \bibinfo {author}
  {\bibfnamefont {C.}~\bibnamefont {Sun}}, \bibinfo {author} {\bibfnamefont
  {F.}~\bibnamefont {Nori}},  \emph {et~al.},\ }\href@noop {} {\bibfield
  {journal} {\bibinfo  {journal} {Physical Review A}\ }\textbf {\bibinfo
  {volume} {80}},\ \bibinfo {pages} {062109} (\bibinfo {year}
  {2009})}\BibitemShut {NoStop}%
\bibitem [{\citenamefont {Hendrickson}\ \emph {et~al.}(2013)\citenamefont
  {Hendrickson}, \citenamefont {Weiler}, \citenamefont {Camacho}, \citenamefont
  {Rakich}, \citenamefont {Young}, \citenamefont {Shaw}, \citenamefont
  {Pittman}, \citenamefont {Franson},\ and\ \citenamefont
  {Jacobs}}]{hendrickson2013all}%
  \BibitemOpen
  \bibfield  {author} {\bibinfo {author} {\bibfnamefont {S.}~\bibnamefont
  {Hendrickson}}, \bibinfo {author} {\bibfnamefont {C.}~\bibnamefont {Weiler}},
  \bibinfo {author} {\bibfnamefont {R.}~\bibnamefont {Camacho}}, \bibinfo
  {author} {\bibfnamefont {P.}~\bibnamefont {Rakich}}, \bibinfo {author}
  {\bibfnamefont {A.}~\bibnamefont {Young}}, \bibinfo {author} {\bibfnamefont
  {M.}~\bibnamefont {Shaw}}, \bibinfo {author} {\bibfnamefont {T.}~\bibnamefont
  {Pittman}}, \bibinfo {author} {\bibfnamefont {J.}~\bibnamefont {Franson}}, \
  and\ \bibinfo {author} {\bibfnamefont {B.}~\bibnamefont {Jacobs}},\
  }\href@noop {} {\bibfield  {journal} {\bibinfo  {journal} {Physical Review
  A}\ }\textbf {\bibinfo {volume} {87}},\ \bibinfo {pages} {023808} (\bibinfo
  {year} {2013})}\BibitemShut {NoStop}%
\bibitem [{\citenamefont {Nodurft}\ \emph
  {et~al.}(2022{\natexlab{c}})\citenamefont {Nodurft}, \citenamefont {Kirby},
  \citenamefont {Glasser}, \citenamefont {Shaw},\ and\ \citenamefont
  {Searles}}]{nodurft2022entangling}%
  \BibitemOpen
  \bibfield  {author} {\bibinfo {author} {\bibfnamefont {I.}~\bibnamefont
  {Nodurft}}, \bibinfo {author} {\bibfnamefont {B.}~\bibnamefont {Kirby}},
  \bibinfo {author} {\bibfnamefont {R.}~\bibnamefont {Glasser}}, \bibinfo
  {author} {\bibfnamefont {H.}~\bibnamefont {Shaw}}, \ and\ \bibinfo {author}
  {\bibfnamefont {T.}~\bibnamefont {Searles}},\ }in\ \href@noop {} {\emph
  {\bibinfo {booktitle} {APS Division of Atomic, Molecular and Optical Physics
  Meeting Abstracts}}},\ Vol.\ \bibinfo {volume} {2022}\ (\bibinfo {year}
  {2022})\ pp.\ \bibinfo {pages} {H06--009}\BibitemShut {NoStop}%
\bibitem [{\citenamefont {ten Brinke}\ \emph {et~al.}(2011)\citenamefont {ten
  Brinke}, \citenamefont {Osterloh},\ and\ \citenamefont
  {Sch{\"u}tzhold}}]{ten2011entangling}%
  \BibitemOpen
  \bibfield  {author} {\bibinfo {author} {\bibfnamefont {N.}~\bibnamefont {ten
  Brinke}}, \bibinfo {author} {\bibfnamefont {A.}~\bibnamefont {Osterloh}}, \
  and\ \bibinfo {author} {\bibfnamefont {R.}~\bibnamefont {Sch{\"u}tzhold}},\
  }\href@noop {} {\bibfield  {journal} {\bibinfo  {journal} {Physical Review
  A}\ }\textbf {\bibinfo {volume} {84}},\ \bibinfo {pages} {022317} (\bibinfo
  {year} {2011})}\BibitemShut {NoStop}%
\bibitem [{\citenamefont {Bayrakci}\ and\ \citenamefont
  {Ozaydin}(2022)}]{bayrakci2022quantum}%
  \BibitemOpen
  \bibfield  {author} {\bibinfo {author} {\bibfnamefont {V.}~\bibnamefont
  {Bayrakci}}\ and\ \bibinfo {author} {\bibfnamefont {F.}~\bibnamefont
  {Ozaydin}},\ }\href@noop {} {\bibfield  {journal} {\bibinfo  {journal}
  {Scientific Reports}\ }\textbf {\bibinfo {volume} {12}},\ \bibinfo {pages}
  {15302} (\bibinfo {year} {2022})}\BibitemShut {NoStop}%
\bibitem [{\citenamefont {Tavakoli}\ \emph {et~al.}(2015)\citenamefont
  {Tavakoli}, \citenamefont {Anwer}, \citenamefont {Hameedi},\ and\
  \citenamefont {Bourennane}}]{tavakoli2015quantum}%
  \BibitemOpen
  \bibfield  {author} {\bibinfo {author} {\bibfnamefont {A.}~\bibnamefont
  {Tavakoli}}, \bibinfo {author} {\bibfnamefont {H.}~\bibnamefont {Anwer}},
  \bibinfo {author} {\bibfnamefont {A.}~\bibnamefont {Hameedi}}, \ and\
  \bibinfo {author} {\bibfnamefont {M.}~\bibnamefont {Bourennane}},\
  }\href@noop {} {\bibfield  {journal} {\bibinfo  {journal} {Physical Review
  A}\ }\textbf {\bibinfo {volume} {92}},\ \bibinfo {pages} {012303} (\bibinfo
  {year} {2015})}\BibitemShut {NoStop}%
\bibitem [{\citenamefont {Cao}\ \emph {et~al.}(2017)\citenamefont {Cao},
  \citenamefont {Li}, \citenamefont {Cao}, \citenamefont {Yin}, \citenamefont
  {Chen}, \citenamefont {Yin}, \citenamefont {Chen}, \citenamefont {Ma},
  \citenamefont {Peng},\ and\ \citenamefont {Pan}}]{cao2017direct}%
  \BibitemOpen
  \bibfield  {author} {\bibinfo {author} {\bibfnamefont {Y.}~\bibnamefont
  {Cao}}, \bibinfo {author} {\bibfnamefont {Y.-H.}\ \bibnamefont {Li}},
  \bibinfo {author} {\bibfnamefont {Z.}~\bibnamefont {Cao}}, \bibinfo {author}
  {\bibfnamefont {J.}~\bibnamefont {Yin}}, \bibinfo {author} {\bibfnamefont
  {Y.-A.}\ \bibnamefont {Chen}}, \bibinfo {author} {\bibfnamefont {H.-L.}\
  \bibnamefont {Yin}}, \bibinfo {author} {\bibfnamefont {T.-Y.}\ \bibnamefont
  {Chen}}, \bibinfo {author} {\bibfnamefont {X.}~\bibnamefont {Ma}}, \bibinfo
  {author} {\bibfnamefont {C.-Z.}\ \bibnamefont {Peng}}, \ and\ \bibinfo
  {author} {\bibfnamefont {J.-W.}\ \bibnamefont {Pan}},\ }\href@noop {}
  {\bibfield  {journal} {\bibinfo  {journal} {Proceedings of the National
  Academy of Sciences}\ }\textbf {\bibinfo {volume} {114}},\ \bibinfo {pages}
  {4920} (\bibinfo {year} {2017})}\BibitemShut {NoStop}%
\bibitem [{\citenamefont {Salih}\ \emph {et~al.}(2022)\citenamefont {Salih},
  \citenamefont {McCutcheon}, \citenamefont {Hance},\ and\ \citenamefont
  {Rarity}}]{salih2022laws}%
  \BibitemOpen
  \bibfield  {author} {\bibinfo {author} {\bibfnamefont {H.}~\bibnamefont
  {Salih}}, \bibinfo {author} {\bibfnamefont {W.}~\bibnamefont {McCutcheon}},
  \bibinfo {author} {\bibfnamefont {J.~R.}\ \bibnamefont {Hance}}, \ and\
  \bibinfo {author} {\bibfnamefont {J.}~\bibnamefont {Rarity}},\ }\href@noop {}
  {\bibfield  {journal} {\bibinfo  {journal} {npj Quantum Information}\
  }\textbf {\bibinfo {volume} {8}},\ \bibinfo {pages} {60} (\bibinfo {year}
  {2022})}\BibitemShut {NoStop}%
\bibitem [{\citenamefont {Salih}\ \emph {et~al.}(2013)\citenamefont {Salih},
  \citenamefont {Li}, \citenamefont {Al-Amri},\ and\ \citenamefont
  {Zubairy}}]{salih2013protocol}%
  \BibitemOpen
  \bibfield  {author} {\bibinfo {author} {\bibfnamefont {H.}~\bibnamefont
  {Salih}}, \bibinfo {author} {\bibfnamefont {Z.-H.}\ \bibnamefont {Li}},
  \bibinfo {author} {\bibfnamefont {M.}~\bibnamefont {Al-Amri}}, \ and\
  \bibinfo {author} {\bibfnamefont {M.~S.}\ \bibnamefont {Zubairy}},\
  }\href@noop {} {\bibfield  {journal} {\bibinfo  {journal} {Physical review
  letters}\ }\textbf {\bibinfo {volume} {110}},\ \bibinfo {pages} {170502}
  (\bibinfo {year} {2013})}\BibitemShut {NoStop}%
\bibitem [{\citenamefont {Vaidman}(2019)}]{vaidman2019analysis}%
  \BibitemOpen
  \bibfield  {author} {\bibinfo {author} {\bibfnamefont {L.}~\bibnamefont
  {Vaidman}},\ }\href@noop {} {\bibfield  {journal} {\bibinfo  {journal}
  {Physical Review A}\ }\textbf {\bibinfo {volume} {99}},\ \bibinfo {pages}
  {052127} (\bibinfo {year} {2019})}\BibitemShut {NoStop}%
\bibitem [{\citenamefont {Cao}\ \emph {et~al.}(2014)\citenamefont {Cao},
  \citenamefont {Li}, \citenamefont {Cao}, \citenamefont {Yin}, \citenamefont
  {Chen}, \citenamefont {Ma}, \citenamefont {Peng},\ and\ \citenamefont
  {Pan}}]{cao2014direct}%
  \BibitemOpen
  \bibfield  {author} {\bibinfo {author} {\bibfnamefont {Y.}~\bibnamefont
  {Cao}}, \bibinfo {author} {\bibfnamefont {Y.-H.}\ \bibnamefont {Li}},
  \bibinfo {author} {\bibfnamefont {Z.}~\bibnamefont {Cao}}, \bibinfo {author}
  {\bibfnamefont {J.}~\bibnamefont {Yin}}, \bibinfo {author} {\bibfnamefont
  {Y.-A.}\ \bibnamefont {Chen}}, \bibinfo {author} {\bibfnamefont
  {X.}~\bibnamefont {Ma}}, \bibinfo {author} {\bibfnamefont {C.-Z.}\
  \bibnamefont {Peng}}, \ and\ \bibinfo {author} {\bibfnamefont {J.-W.}\
  \bibnamefont {Pan}},\ }in\ \href@noop {} {\emph {\bibinfo {booktitle} {CLEO:
  QELS\_Fundamental Science}}}\ (\bibinfo {organization} {Optica Publishing
  Group},\ \bibinfo {year} {2014})\ pp.\ \bibinfo {pages} {FM4A--6}\BibitemShut
  {NoStop}%
\bibitem [{\citenamefont {Alonso~Calafell}\ \emph {et~al.}(2019)\citenamefont
  {Alonso~Calafell}, \citenamefont {Str{\"o}mberg}, \citenamefont
  {Arvidsson-Shukur}, \citenamefont {Rozema}, \citenamefont {Saggio},
  \citenamefont {Greganti}, \citenamefont {Harris}, \citenamefont {Prabhu},
  \citenamefont {Carolan}, \citenamefont {Hochberg} \emph
  {et~al.}}]{alonso2019trace}%
  \BibitemOpen
  \bibfield  {author} {\bibinfo {author} {\bibfnamefont {I.}~\bibnamefont
  {Alonso~Calafell}}, \bibinfo {author} {\bibfnamefont {T.}~\bibnamefont
  {Str{\"o}mberg}}, \bibinfo {author} {\bibfnamefont {D.}~\bibnamefont
  {Arvidsson-Shukur}}, \bibinfo {author} {\bibfnamefont {L.}~\bibnamefont
  {Rozema}}, \bibinfo {author} {\bibfnamefont {V.}~\bibnamefont {Saggio}},
  \bibinfo {author} {\bibfnamefont {C.}~\bibnamefont {Greganti}}, \bibinfo
  {author} {\bibfnamefont {N.}~\bibnamefont {Harris}}, \bibinfo {author}
  {\bibfnamefont {M.}~\bibnamefont {Prabhu}}, \bibinfo {author} {\bibfnamefont
  {J.}~\bibnamefont {Carolan}}, \bibinfo {author} {\bibfnamefont
  {M.}~\bibnamefont {Hochberg}},  \emph {et~al.},\ }\href@noop {} {\bibfield
  {journal} {\bibinfo  {journal} {npj Quantum Information}\ }\textbf {\bibinfo
  {volume} {5}},\ \bibinfo {pages} {61} (\bibinfo {year} {2019})}\BibitemShut
  {NoStop}%
\bibitem [{\citenamefont {Li}\ \emph {et~al.}(2015)\citenamefont {Li},
  \citenamefont {Al-Amri},\ and\ \citenamefont {Zubairy}}]{li2015direct}%
  \BibitemOpen
  \bibfield  {author} {\bibinfo {author} {\bibfnamefont {Z.-H.}\ \bibnamefont
  {Li}}, \bibinfo {author} {\bibfnamefont {M.}~\bibnamefont {Al-Amri}}, \ and\
  \bibinfo {author} {\bibfnamefont {M.~S.}\ \bibnamefont {Zubairy}},\
  }\href@noop {} {\bibfield  {journal} {\bibinfo  {journal} {Physical Review
  A}\ }\textbf {\bibinfo {volume} {92}},\ \bibinfo {pages} {052315} (\bibinfo
  {year} {2015})}\BibitemShut {NoStop}%
\bibitem [{\citenamefont {Zaman}\ \emph {et~al.}(2018)\citenamefont {Zaman},
  \citenamefont {Jeong},\ and\ \citenamefont {Shin}}]{zaman2018counterfactual}%
  \BibitemOpen
  \bibfield  {author} {\bibinfo {author} {\bibfnamefont {F.}~\bibnamefont
  {Zaman}}, \bibinfo {author} {\bibfnamefont {Y.}~\bibnamefont {Jeong}}, \ and\
  \bibinfo {author} {\bibfnamefont {H.}~\bibnamefont {Shin}},\ }\href@noop {}
  {\bibfield  {journal} {\bibinfo  {journal} {Scientific reports}\ }\textbf
  {\bibinfo {volume} {8}},\ \bibinfo {pages} {14641} (\bibinfo {year}
  {2018})}\BibitemShut {NoStop}%
\bibitem [{\citenamefont {Ma}\ \emph {et~al.}(2014)\citenamefont {Ma},
  \citenamefont {Guo}, \citenamefont {Schuck}, \citenamefont {Fong},
  \citenamefont {Jiang},\ and\ \citenamefont {Tang}}]{ma2014chip}%
  \BibitemOpen
  \bibfield  {author} {\bibinfo {author} {\bibfnamefont {X.-s.}\ \bibnamefont
  {Ma}}, \bibinfo {author} {\bibfnamefont {X.}~\bibnamefont {Guo}}, \bibinfo
  {author} {\bibfnamefont {C.}~\bibnamefont {Schuck}}, \bibinfo {author}
  {\bibfnamefont {K.~Y.}\ \bibnamefont {Fong}}, \bibinfo {author}
  {\bibfnamefont {L.}~\bibnamefont {Jiang}}, \ and\ \bibinfo {author}
  {\bibfnamefont {H.~X.}\ \bibnamefont {Tang}},\ }\href@noop {} {\bibfield
  {journal} {\bibinfo  {journal} {Physical Review A}\ }\textbf {\bibinfo
  {volume} {90}},\ \bibinfo {pages} {042109} (\bibinfo {year}
  {2014})}\BibitemShut {NoStop}%
\bibitem [{\citenamefont {Leppenen}\ \emph {et~al.}(2021)\citenamefont
  {Leppenen}, \citenamefont {Lanco},\ and\ \citenamefont
  {Smirnov}}]{leppenen2021quantum}%
  \BibitemOpen
  \bibfield  {author} {\bibinfo {author} {\bibfnamefont {N.}~\bibnamefont
  {Leppenen}}, \bibinfo {author} {\bibfnamefont {L.}~\bibnamefont {Lanco}}, \
  and\ \bibinfo {author} {\bibfnamefont {D.}~\bibnamefont {Smirnov}},\
  }\href@noop {} {\bibfield  {journal} {\bibinfo  {journal} {Physical Review
  B}\ }\textbf {\bibinfo {volume} {103}},\ \bibinfo {pages} {045413} (\bibinfo
  {year} {2021})}\BibitemShut {NoStop}%
\bibitem [{\citenamefont {Helmer}\ \emph {et~al.}(2007)\citenamefont {Helmer},
  \citenamefont {Mariantoni}, \citenamefont {Solano},\ and\ \citenamefont
  {Marquardt}}]{helmer2007quantum}%
  \BibitemOpen
  \bibfield  {author} {\bibinfo {author} {\bibfnamefont {F.}~\bibnamefont
  {Helmer}}, \bibinfo {author} {\bibfnamefont {M.}~\bibnamefont {Mariantoni}},
  \bibinfo {author} {\bibfnamefont {E.}~\bibnamefont {Solano}}, \ and\ \bibinfo
  {author} {\bibfnamefont {F.}~\bibnamefont {Marquardt}},\ }\href@noop {}
  {\bibfield  {journal} {\bibinfo  {journal} {arXiv preprint arXiv:0712.1908}\
  } (\bibinfo {year} {2007})}\BibitemShut {NoStop}%
\bibitem [{\citenamefont {Maniscalco}\ \emph {et~al.}(2008)\citenamefont
  {Maniscalco}, \citenamefont {Francica}, \citenamefont {Zaffino},
  \citenamefont {Gullo},\ and\ \citenamefont
  {Plastina}}]{maniscalco2008protecting}%
  \BibitemOpen
  \bibfield  {author} {\bibinfo {author} {\bibfnamefont {S.}~\bibnamefont
  {Maniscalco}}, \bibinfo {author} {\bibfnamefont {F.}~\bibnamefont
  {Francica}}, \bibinfo {author} {\bibfnamefont {R.~L.}\ \bibnamefont
  {Zaffino}}, \bibinfo {author} {\bibfnamefont {N.~L.}\ \bibnamefont {Gullo}},
  \ and\ \bibinfo {author} {\bibfnamefont {F.}~\bibnamefont {Plastina}},\
  }\href@noop {} {\bibfield  {journal} {\bibinfo  {journal} {Physical review
  letters}\ }\textbf {\bibinfo {volume} {100}},\ \bibinfo {pages} {090503}
  (\bibinfo {year} {2008})}\BibitemShut {NoStop}%
\bibitem [{\citenamefont {Hacohen-Gourgy}\ \emph {et~al.}(2018)\citenamefont
  {Hacohen-Gourgy}, \citenamefont {Garc{\'{\i}}a-Pintos}, \citenamefont
  {Martin}, \citenamefont {Dressel},\ and\ \citenamefont
  {Siddiqi}}]{HacohenGourgy2018}%
  \BibitemOpen
  \bibfield  {author} {\bibinfo {author} {\bibfnamefont {S.}~\bibnamefont
  {Hacohen-Gourgy}}, \bibinfo {author} {\bibfnamefont {L.}~\bibnamefont
  {Garc{\'{\i}}a-Pintos}}, \bibinfo {author} {\bibfnamefont {L.}~\bibnamefont
  {Martin}}, \bibinfo {author} {\bibfnamefont {J.}~\bibnamefont {Dressel}}, \
  and\ \bibinfo {author} {\bibfnamefont {I.}~\bibnamefont {Siddiqi}},\ }\href
  {\doibase 10.1103/physrevlett.120.020505} {\bibfield  {journal} {\bibinfo
  {journal} {Physical Review Letters}\ }\textbf {\bibinfo {volume} {120}}
  (\bibinfo {year} {2018}),\ 10.1103/physrevlett.120.020505}\BibitemShut
  {NoStop}%
\bibitem [{\citenamefont {Paz-Silva}\ \emph {et~al.}(2012)\citenamefont
  {Paz-Silva}, \citenamefont {Rezakhani}, \citenamefont {Dominy},\ and\
  \citenamefont {Lidar}}]{PazSilva2012}%
  \BibitemOpen
  \bibfield  {author} {\bibinfo {author} {\bibfnamefont {G.~A.}\ \bibnamefont
  {Paz-Silva}}, \bibinfo {author} {\bibfnamefont {A.~T.}\ \bibnamefont
  {Rezakhani}}, \bibinfo {author} {\bibfnamefont {J.~M.}\ \bibnamefont
  {Dominy}}, \ and\ \bibinfo {author} {\bibfnamefont {D.~A.}\ \bibnamefont
  {Lidar}},\ }\href {\doibase 10.1103/physrevlett.108.080501} {\bibfield
  {journal} {\bibinfo  {journal} {Physical Review Letters}\ }\textbf {\bibinfo
  {volume} {108}} (\bibinfo {year} {2012}),\
  10.1103/physrevlett.108.080501}\BibitemShut {NoStop}%
\bibitem [{\citenamefont {Wang}\ \emph {et~al.}(2008)\citenamefont {Wang},
  \citenamefont {You},\ and\ \citenamefont {Nori}}]{Wang2008}%
  \BibitemOpen
  \bibfield  {author} {\bibinfo {author} {\bibfnamefont {X.-B.}\ \bibnamefont
  {Wang}}, \bibinfo {author} {\bibfnamefont {J.~Q.}\ \bibnamefont {You}}, \
  and\ \bibinfo {author} {\bibfnamefont {F.}~\bibnamefont {Nori}},\ }\href
  {\doibase 10.1103/physreva.77.062339} {\bibfield  {journal} {\bibinfo
  {journal} {Physical Review A}\ }\textbf {\bibinfo {volume} {77}} (\bibinfo
  {year} {2008}),\ 10.1103/physreva.77.062339}\BibitemShut {NoStop}%
\bibitem [{\citenamefont {Kofman}\ and\ \citenamefont
  {Kurizki}(2004)}]{Kofman2004}%
  \BibitemOpen
  \bibfield  {author} {\bibinfo {author} {\bibfnamefont {A.~G.}\ \bibnamefont
  {Kofman}}\ and\ \bibinfo {author} {\bibfnamefont {G.}~\bibnamefont
  {Kurizki}},\ }\href {\doibase 10.1103/physrevlett.93.130406} {\bibfield
  {journal} {\bibinfo  {journal} {Physical Review Letters}\ }\textbf {\bibinfo
  {volume} {93}} (\bibinfo {year} {2004}),\
  10.1103/physrevlett.93.130406}\BibitemShut {NoStop}%
\bibitem [{\citenamefont {Harrington}\ \emph {et~al.}(2017)\citenamefont
  {Harrington}, \citenamefont {Monroe},\ and\ \citenamefont
  {Murch}}]{Harrington2017}%
  \BibitemOpen
  \bibfield  {author} {\bibinfo {author} {\bibfnamefont {P.}~\bibnamefont
  {Harrington}}, \bibinfo {author} {\bibfnamefont {J.}~\bibnamefont {Monroe}},
  \ and\ \bibinfo {author} {\bibfnamefont {K.}~\bibnamefont {Murch}},\ }\href
  {\doibase 10.1103/physrevlett.118.240401} {\bibfield  {journal} {\bibinfo
  {journal} {Physical Review Letters}\ }\textbf {\bibinfo {volume} {118}}
  (\bibinfo {year} {2017}),\ 10.1103/physrevlett.118.240401}\BibitemShut
  {NoStop}%
\bibitem [{\citenamefont {Franson}\ \emph {et~al.}(2004)\citenamefont
  {Franson}, \citenamefont {Jacobs},\ and\ \citenamefont
  {Pittman}}]{franson2004quantum}%
  \BibitemOpen
  \bibfield  {author} {\bibinfo {author} {\bibfnamefont {J.~D.}\ \bibnamefont
  {Franson}}, \bibinfo {author} {\bibfnamefont {B.~C.}\ \bibnamefont {Jacobs}},
  \ and\ \bibinfo {author} {\bibfnamefont {T.~B.}\ \bibnamefont {Pittman}},\
  }\href@noop {} {\bibfield  {journal} {\bibinfo  {journal} {Physical Review
  A}\ }\textbf {\bibinfo {volume} {70}},\ \bibinfo {pages} {062302} (\bibinfo
  {year} {2004})}\BibitemShut {NoStop}%
\bibitem [{\citenamefont {M\"{u}ller}\ \emph {et~al.}(2017)\citenamefont
  {M\"{u}ller}, \citenamefont {Gherardini},\ and\ \citenamefont
  {Caruso}}]{Mller2017}%
  \BibitemOpen
  \bibfield  {author} {\bibinfo {author} {\bibfnamefont {M.~M.}\ \bibnamefont
  {M\"{u}ller}}, \bibinfo {author} {\bibfnamefont {S.}~\bibnamefont
  {Gherardini}}, \ and\ \bibinfo {author} {\bibfnamefont {F.}~\bibnamefont
  {Caruso}},\ }\href {\doibase 10.1002/andp.201600206} {\bibfield  {journal}
  {\bibinfo  {journal} {Annalen der Physik}\ }\textbf {\bibinfo {volume}
  {529}},\ \bibinfo {pages} {1600206} (\bibinfo {year} {2017})}\BibitemShut
  {NoStop}%
\bibitem [{\citenamefont {Kofman}\ and\ \citenamefont
  {Kurizki}(2001)}]{Kofman2001}%
  \BibitemOpen
  \bibfield  {author} {\bibinfo {author} {\bibfnamefont {A.~G.}\ \bibnamefont
  {Kofman}}\ and\ \bibinfo {author} {\bibfnamefont {G.}~\bibnamefont
  {Kurizki}},\ }\href {\doibase 10.1103/physrevlett.87.270405} {\bibfield
  {journal} {\bibinfo  {journal} {Physical Review Letters}\ }\textbf {\bibinfo
  {volume} {87}} (\bibinfo {year} {2001}),\
  10.1103/physrevlett.87.270405}\BibitemShut {NoStop}%
\bibitem [{\citenamefont {Pechen}\ \emph {et~al.}(2006)\citenamefont {Pechen},
  \citenamefont {Il'in}, \citenamefont {Shuang},\ and\ \citenamefont
  {Rabitz}}]{Pechen2006}%
  \BibitemOpen
  \bibfield  {author} {\bibinfo {author} {\bibfnamefont {A.}~\bibnamefont
  {Pechen}}, \bibinfo {author} {\bibfnamefont {N.}~\bibnamefont {Il'in}},
  \bibinfo {author} {\bibfnamefont {F.}~\bibnamefont {Shuang}}, \ and\ \bibinfo
  {author} {\bibfnamefont {H.}~\bibnamefont {Rabitz}},\ }\href {\doibase
  10.1103/physreva.74.052102} {\bibfield  {journal} {\bibinfo  {journal}
  {Physical Review A}\ }\textbf {\bibinfo {volume} {74}} (\bibinfo {year}
  {2006}),\ 10.1103/physreva.74.052102}\BibitemShut {NoStop}%
\bibitem [{\citenamefont {S{\o}rensen}\ \emph {et~al.}(2018)\citenamefont
  {S{\o}rensen}, \citenamefont {Dalgaard}, \citenamefont {Kiilerich},
  \citenamefont {M{\o}lmer},\ and\ \citenamefont {Sherson}}]{Srensen2018}%
  \BibitemOpen
  \bibfield  {author} {\bibinfo {author} {\bibfnamefont {J.~J. W.~H.}\
  \bibnamefont {S{\o}rensen}}, \bibinfo {author} {\bibfnamefont
  {M.}~\bibnamefont {Dalgaard}}, \bibinfo {author} {\bibfnamefont {A.~H.}\
  \bibnamefont {Kiilerich}}, \bibinfo {author} {\bibfnamefont {K.}~\bibnamefont
  {M{\o}lmer}}, \ and\ \bibinfo {author} {\bibfnamefont {J.~F.}\ \bibnamefont
  {Sherson}},\ }\href {\doibase 10.1103/physreva.98.062317} {\bibfield
  {journal} {\bibinfo  {journal} {Physical Review A}\ }\textbf {\bibinfo
  {volume} {98}} (\bibinfo {year} {2018}),\
  10.1103/physreva.98.062317}\BibitemShut {NoStop}%
\bibitem [{\citenamefont {Piacentini}\ \emph {et~al.}(2017)\citenamefont
  {Piacentini}, \citenamefont {Avella}, \citenamefont {Rebufello},
  \citenamefont {Lussana}, \citenamefont {Villa}, \citenamefont {Tosi},
  \citenamefont {Gramegna}, \citenamefont {Brida}, \citenamefont {Cohen},
  \citenamefont {Vaidman}, \citenamefont {Degiovanni},\ and\ \citenamefont
  {Genovese}}]{Piacentini2017}%
  \BibitemOpen
  \bibfield  {author} {\bibinfo {author} {\bibfnamefont {F.}~\bibnamefont
  {Piacentini}}, \bibinfo {author} {\bibfnamefont {A.}~\bibnamefont {Avella}},
  \bibinfo {author} {\bibfnamefont {E.}~\bibnamefont {Rebufello}}, \bibinfo
  {author} {\bibfnamefont {R.}~\bibnamefont {Lussana}}, \bibinfo {author}
  {\bibfnamefont {F.}~\bibnamefont {Villa}}, \bibinfo {author} {\bibfnamefont
  {A.}~\bibnamefont {Tosi}}, \bibinfo {author} {\bibfnamefont {M.}~\bibnamefont
  {Gramegna}}, \bibinfo {author} {\bibfnamefont {G.}~\bibnamefont {Brida}},
  \bibinfo {author} {\bibfnamefont {E.}~\bibnamefont {Cohen}}, \bibinfo
  {author} {\bibfnamefont {L.}~\bibnamefont {Vaidman}}, \bibinfo {author}
  {\bibfnamefont {I.~P.}\ \bibnamefont {Degiovanni}}, \ and\ \bibinfo {author}
  {\bibfnamefont {M.}~\bibnamefont {Genovese}},\ }\href {\doibase
  10.1038/nphys4223} {\bibfield  {journal} {\bibinfo  {journal} {Nature
  Physics}\ }\textbf {\bibinfo {volume} {13}},\ \bibinfo {pages} {1191}
  (\bibinfo {year} {2017})}\BibitemShut {NoStop}%
\bibitem [{\citenamefont {Barontini}\ \emph {et~al.}(2015)\citenamefont
  {Barontini}, \citenamefont {Hohmann}, \citenamefont {Haas}, \citenamefont
  {Est{\`{e}}ve},\ and\ \citenamefont {Reichel}}]{Barontini2015}%
  \BibitemOpen
  \bibfield  {author} {\bibinfo {author} {\bibfnamefont {G.}~\bibnamefont
  {Barontini}}, \bibinfo {author} {\bibfnamefont {L.}~\bibnamefont {Hohmann}},
  \bibinfo {author} {\bibfnamefont {F.}~\bibnamefont {Haas}}, \bibinfo {author}
  {\bibfnamefont {J.}~\bibnamefont {Est{\`{e}}ve}}, \ and\ \bibinfo {author}
  {\bibfnamefont {J.}~\bibnamefont {Reichel}},\ }\href {\doibase
  10.1126/science.aaa0754} {\bibfield  {journal} {\bibinfo  {journal}
  {Science}\ }\textbf {\bibinfo {volume} {349}},\ \bibinfo {pages} {1317}
  (\bibinfo {year} {2015})}\BibitemShut {NoStop}%
\bibitem [{\citenamefont {Kofman}\ \emph {et~al.}(2001)\citenamefont {Kofman},
  \citenamefont {Kurizki},\ and\ \citenamefont {Opatrn{\'{y}}}}]{Kofman2001b}%
  \BibitemOpen
  \bibfield  {author} {\bibinfo {author} {\bibfnamefont {A.~G.}\ \bibnamefont
  {Kofman}}, \bibinfo {author} {\bibfnamefont {G.}~\bibnamefont {Kurizki}}, \
  and\ \bibinfo {author} {\bibfnamefont {T.}~\bibnamefont {Opatrn{\'{y}}}},\
  }\href {\doibase 10.1103/physreva.63.042108} {\bibfield  {journal} {\bibinfo
  {journal} {Physical Review A}\ }\textbf {\bibinfo {volume} {63}} (\bibinfo
  {year} {2001}),\ 10.1103/physreva.63.042108}\BibitemShut {NoStop}%
\bibitem [{\citenamefont {Clausen}\ \emph {et~al.}(2010)\citenamefont
  {Clausen}, \citenamefont {Bensky},\ and\ \citenamefont
  {Kurizki}}]{Clausen2010}%
  \BibitemOpen
  \bibfield  {author} {\bibinfo {author} {\bibfnamefont {J.}~\bibnamefont
  {Clausen}}, \bibinfo {author} {\bibfnamefont {G.}~\bibnamefont {Bensky}}, \
  and\ \bibinfo {author} {\bibfnamefont {G.}~\bibnamefont {Kurizki}},\ }\href
  {\doibase 10.1103/physrevlett.104.040401} {\bibfield  {journal} {\bibinfo
  {journal} {Physical Review Letters}\ }\textbf {\bibinfo {volume} {104}}
  (\bibinfo {year} {2010}),\ 10.1103/physrevlett.104.040401}\BibitemShut
  {NoStop}%
\bibitem [{\citenamefont {Zhang}\ \emph {et~al.}(2018)\citenamefont {Zhang},
  \citenamefont {Jing}, \citenamefont {Wang},\ and\ \citenamefont
  {Zhu}}]{PhysRevA.98.012135}%
  \BibitemOpen
  \bibfield  {author} {\bibinfo {author} {\bibfnamefont {J.-M.}\ \bibnamefont
  {Zhang}}, \bibinfo {author} {\bibfnamefont {J.}~\bibnamefont {Jing}},
  \bibinfo {author} {\bibfnamefont {L.-G.}\ \bibnamefont {Wang}}, \ and\
  \bibinfo {author} {\bibfnamefont {S.-Y.}\ \bibnamefont {Zhu}},\ }\href
  {\doibase 10.1103/PhysRevA.98.012135} {\bibfield  {journal} {\bibinfo
  {journal} {Phys. Rev. A}\ }\textbf {\bibinfo {volume} {98}},\ \bibinfo
  {pages} {012135} (\bibinfo {year} {2018})}\BibitemShut {NoStop}%
\bibitem [{\citenamefont {von Neumann}(1955)}]{vonNeumann1955}%
  \BibitemOpen
  \bibfield  {author} {\bibinfo {author} {\bibfnamefont {J.}~\bibnamefont {von
  Neumann}},\ }\href@noop {} {\emph {\bibinfo {title} {Mathematical Foundations
  of Quantum Mechanics}}}\ (\bibinfo  {publisher} {Princeton University
  Press},\ \bibinfo {address} {Princeton},\ \bibinfo {year} {1955})\BibitemShut
  {NoStop}%
\bibitem [{\citenamefont {Saha}\ and\ \citenamefont
  {Batista}(2011)}]{saha2011tunneling}%
  \BibitemOpen
  \bibfield  {author} {\bibinfo {author} {\bibfnamefont {R.}~\bibnamefont
  {Saha}}\ and\ \bibinfo {author} {\bibfnamefont {V.~S.}\ \bibnamefont
  {Batista}},\ }\href@noop {} {\bibfield  {journal} {\bibinfo  {journal} {The
  Journal of Physical Chemistry B}\ }\textbf {\bibinfo {volume} {115}},\
  \bibinfo {pages} {5234} (\bibinfo {year} {2011})}\BibitemShut {NoStop}%
\bibitem [{\citenamefont {Saha}\ \emph {et~al.}(2012)\citenamefont {Saha},
  \citenamefont {Markmann},\ and\ \citenamefont {Batista}}]{saha2012tunneling}%
  \BibitemOpen
  \bibfield  {author} {\bibinfo {author} {\bibfnamefont {R.}~\bibnamefont
  {Saha}}, \bibinfo {author} {\bibfnamefont {A.}~\bibnamefont {Markmann}}, \
  and\ \bibinfo {author} {\bibfnamefont {V.~S.}\ \bibnamefont {Batista}},\
  }\href@noop {} {\bibfield  {journal} {\bibinfo  {journal} {Molecular
  Physics}\ }\textbf {\bibinfo {volume} {110}},\ \bibinfo {pages} {995}
  (\bibinfo {year} {2012})}\BibitemShut {NoStop}%
\end{thebibliography}

\end{document}